\documentclass[twoside,11pt]{article}

\RequirePackage{natbib}
\usepackage{hyperref}
\usepackage{xcolor}
\usepackage{amsmath}
\usepackage{setspace}
\usepackage{amsthm}
\usepackage{authblk}
\usepackage{amssymb}
\usepackage{multirow}
\usepackage{adjustbox}
\usepackage{rotating}
\usepackage[margin = 1.25in]{geometry}
\usepackage[toc, page]{appendix}

\if@abbrvbib
\else
\fi

\bibpunct{(}{)}{;}{a}{,}{,}

\newenvironment{keywords}
{\bgroup\leftskip 20pt\rightskip 20pt \small\noindent{\bf Keywords:} }%
{\par\egroup\vskip 0.25ex}

\begin{document}

\title{\textbf{Gaussian Process Mixtures for Estimating Heterogeneous Treatment Effects}}

\author{Abbas Zaidi}
\affil{\textit{Department of Statistical Science, Duke University}.\footnote{abbas.zaidi@duke.edu}}
\author{Sayan Mukherjee}
\affil{\textit{Departments of Statistical Science, Mathematics,
       Computer Science, Biostatistics and Bioinformatics, Duke University}.\footnote{sayan@stat.duke.edu}}

\maketitle
\onehalfspace

\begin{abstract}
We develop a Gaussian-process mixture model for heterogeneous treatment effect estimation that leverages the use of transformed outcomes. The approach we will present attempts to improve point estimation and uncertainty quantification relative to past work that has used transformed variable related methods as well as traditional outcome modeling. Earlier work on modeling treatment effect heterogeneity using transformed outcomes has relied on tree based methods such as single regression trees and random forests. Under the umbrella of non-parametric models, outcome modeling has been performed using Bayesian additive regression trees and various flavors of weighted single trees. These approaches work well when large samples are available, but suffer in smaller samples where results are more sensitive to model misspecification -- our method attempts to garner improvements in inference quality via a correctly specified model rooted in Bayesian non-parametrics. Furthermore, while we begin with a model that assumes that the treatment assignment mechanism is known, an extension where it is learnt from the data is presented for applications to observational studies. Our approach is applied to simulated and real data to demonstrate our theorized improvements in inference with respect to two causal estimands: the conditional average treatment effect and the average treatment effect. By leveraging our correctly specified model, we are able to more accurately estimate the treatment effects while reducing their variance. 
\end{abstract}


\section{Introduction}
The estimation of treatment effects is one of the core problems in causal inference. A treatment effect is a measure used to compare interventions in randomized experiments, policy analysis, and medical trials. The treatment effect measures the difference in outcomes between units assigned to the treatment versus those assigned to the control. There have been a variety of related approaches for estimating treatment effects including those based on graphical models \citep{pearl2009causal} and the potential outcomes framework \citep{rubin1978bayesian}. In this paper, we develop methodology that builds on the potential outcomes framework as defined in  \cite{rubin2005causal} to estimate treatment effects.

In the potential outcomes framework we compare the observed outcome to the outcome under the counterfactual, that is, what the outcome would be under a different set of treatment conditions. If the counterfactual outcome were known then the treatment effect on an individual unit is the difference between the outcome under the observed and counterfactual interventions. The \emph{fundamental problem of causal inference} is that in general for any unit  one can only observe the outcome under a single treatment condition. As a consequence unit level causal effects are not identifiable.  However, population level causal effects can be identified under some standard assumptions (see Section \ref{former}). An estimator of population level effects is the average treatment effect (ATE) which is a measure of the difference in the mean outcomes between units assigned to the treatment and units assigned to the control. If treatment effects are homogenous across individuals then estimators such as the ATE that consider causal effects at an aggregate level are reasonable, however such estimators will overlook subgroup or covariate-level  specific heterogeneity in treatment effects. There is evidence that heterogeneity in treatment effects is more the rule than the exception \citep{heckman2006understanding, green2012modeling, xie2012estimating}.

A quantity in addressing heterogeneous treatment effects is the conditional average treatment effect (CATE) which is the average treatment effect conditional on the covariate level of a unit of observation. One can consider the CATE as a difference of two regression functions -- the average response given treatment at a set of covariate levels minus the the average response assuming the control condition and the same set of covariate levels. One can estimate the ATE by marginalizing the CATE over the joint distribution of the covariates. There are a number of approaches for estimating the two aforementioned causal estimands. The main approach for modeling heterogeneous treatment effects based on the CATE is conditional mean regression. Under this approach, we model the CATE as a difference between the conditional mean outcome given the treatment for particular covariate levels minus the mean outcome given the control at the same covariate levels \citep{ding2017causal}. The implementation of these models can be approached both parametrically and non-parametrically.

The most popular parametric methods for estimating the difference between the conditional mean outcomes include linear and polynomial regression \citep{pearl2009causal}, along with penalized regression approaches such as least absolute subset selection operator and ridge regression \citep{tibshirani1996regression}. At the other end of the spectrum are non-parametric regression models to  estimate the difference between the conditional means. Examples include boosting \citep{powers2017some}, Bayesian additive regression trees (BART) \citep{hill2011bayesian, hahn2017bayesian, chipman2010bart} as well as classical regression trees \citep{athey2015machine, breiman1984classification} and random forests \citep{wager2017estimation, foster2011subgroup, breiman2001random}. These methods have some limitations to their use and we provide a brief discussion of these.

The use of random forests for CATE estimation as defined in \cite{wager2017estimation} provides some interesting theoretical results that allow for probabilistically valid statistical inference. These methods are theorized to outperform classical methods particularly in the presence of irrelevant covariates. This technique however, has been demonstrated to be outperformed in application \citep{hahn2017bayesian}. In addition, without a procedure for imposing a degree of regularization, random forests are difficult to actually deploy for heterogeneous treatment effect estimation \citep{wendling2018comparing}. BART and its variants \citep{hahn2017bayesian, hill2011bayesian} present a persuasive argument for their use in application, but there is limited work on their formal inferential properties \citep{wager2017estimation} for learning heterogeneous treatment effects. Specifically for BART, formal statistical analysis is hurdled by the lack of theory arguing posterior concentration around the true conditional mean function -- the key quantity of interest in heterogeneous treatment effect estimation via conditional mean regression.

An alternative to modeling the difference in conditional mean outcomes is the use of transformed responses or outcome variables (TRV) \citep{dudik2011doubly, beygelzimer2009offset} that is ideologically similar to concepts of \emph{inverse probability weighting} (IPW) \citep{hirano2003efficient}. The TRV approach introduces a transformation for the outcome and the treatment indicator variable for which the conditional expectation given a covariate level is equivalent to the CATE. This allows it to be used with \emph{off-the-shelf} machine learning techniques and has been applied to optimal treatment policy estimation in the same vein as ideas of \emph{double-robustness} as reviewed in \cite{ding2017causal} that combine regression adjustment with weighting. More recent work on the TRV has attempted to model it as a function of the observed covariates via regression trees \citep{athey2015machine} and boosting  \citep{powers2017some}. This has raised questions of estimation quality of the approach given the high variance of the procedure. We assert that this is a consequence of the properties of the TRV that have not been explicitly accounted for in the model since past work has relied on using it as a benchmark for other methods \citep{athey2016recursive}.

In this paper we introduce a novel non-parametric Bayesian model based on Gaussian process regression \citep{singh2016gaussian, rasmussen2006Gaussian} for inference of the TRV that allows us to infer a posterior distribution on the CATE. The model we propose is a finite mixture of Gaussian-processes \citep{rasmussen2000infinite} that leverages the distribution implied by the transformation. This formulation is aimed at improving the overall quality of inference on the treatment effects with a correctly specified model. 

This approach has benefits over both conditional mean regression and other TRV based techniques. In practice, we never estimate either the treatment nor the control function perfectly and different covariate distributions for the treatment and control groups can lead to biases in the treatment effect estimation \citep{powers2017some}. The TRV allows for the joint modeling of information from both the treated and control groups which can help circumvent the aforementioned estimation challenge which for instance has been discussed as a specific limitation of conditional mean regression with random forests \citep{wager2017estimation}. This joint modeling is also an improvement over Bayesian techniques that place individual vague priors on the treatment and control outcome models since the prior on the treatment effect as the difference of the two is possibly \emph{doubly} vague \citep{hahn2017bayesian}. This can make inference a challenge since it is difficult to control the degree of heterogeneity that the model adapts to. Furthermore the TRV generates unbiased estimates for the CATE \citep{powers2017some}.

In addition to its benefits over conditional mean regression methods, the model we introduce offers four advantages over other TRV modeling approaches. First, we significantly improve the accuracy of point estimation by explicitly modeling the distribution of the transformed outcome. Second, by modeling the distribution of the transformed outcome specifically we are able to greatly reduce the variance of causal estimands i.e. the average treatment effect and the conditional average treatment effect. Reducing the variance of the estimators is crucial since this has been the main criticism of the TRV approach  \citep{athey2015machine, powers2017some}. This provides tighter uncertainty intervals relative to the approaches discussed in \cite{athey2015machine} and \cite{wager2017estimation}. Third, our approach is well suited for instances when the treated and control groups share information since our proposed mechanism jointly models the behavior of both via the transformation.

The methodology we introduce makes a number of significant contributions to the estimation of heterogeneous treatment effects. Our main contribution is that we improve the overall quality of inference by improving the point estimation with a correctly specified model. In addition, the proposed framework is flexible in that we do not assume a functional form for how heterogeneity of treatment effects are driven by the levels of the observed covariates. Finally, our proposed framework is easily adapted to studies where the mechanism by which individuals receive the treatment is unknown. For this problem, past work has relied on a two-stage procedure for learning this treatment assignment mechanism first and then utilizing this in the model. We instead propose an approach whereby the treatment assignment mechanism and the treatment effects are jointly learnt in a unified framework. By working under this paradigm we have a twofold gain. First, the uncertainty quantification from our proposed model reflects uncertainty from all stages of inference including the learning of the assignment mechanism, and the treatment effects. Second as a by-product of the \emph{feedback} in between the two estimation stages, the assignment mechanism makes more complete use of the data, which can improve estimation of causal effects.

The remainder of this paper is organized as follows: in Section \ref{former}, we introduce the TRV, the relevant notation and the assumptions inherent to the TRV approach. We state our new model in Section \ref{model}. Our approach is benchmarked against to TRV regression trees and random forests, along with non-TRV weighted tree methods as discussed in \cite{athey2016recursive}, as well as Bayesian tree models in \citep{hahn2017bayesian, hill2011bayesian} on both simulated and real data in Section  \ref{data}. We close with a summary of our findings and possible areas of future work.

\section{Transformed Response Variables Framework}\label{former}
In order to formulate the approach of transformed outcomes, we first define some notation that we will use throughout this paper. The observed data $\mathcal{D}$ consists of a sample of size $n$ where for each unit of observation we are given a response variable $Y_{i} \in \mathbb{R}$ and a covariate vector$X_{i} \in \mathbb{R}^{p}$. In addition to the observed data, we denote as $W_{i} \in \{0, 1\}$ the treatment assignment. The corresponding treatment assignment probability is denoted as $e_{i} = \mathbb{P}(W_{i} = 1)$. Finally, the potential outcome is denoted as  $Y_{i}(W_{i} = w)$. 

Under the potential outcomes framework, in order to estimate treatment effects from observational data certain assumptions about the treatment assignment mechanism need to be satisfied. Briefly, these assumptions are that the treatment assignment is \emph{individualistic} (A1), \emph{probabilistic} (A2) and \emph{ignorable}(A3). Details of these assumptions are left to the reader in \cite{imbens2015causal}. A1 and A2 are implied under the assumption that the units of observation are a simple random sample from the target population that are independent and identically distributed. 

Assumptions A2 and A3 are together known as the  \emph{strong ignorability} assumption and grants the indentifying equivalence between the potential outcome and the causal conditioning,
$Y(W = w) \stackrel{P}{=} Y \mid W = w$. All three of the assumptions summarized here are always satisfied in randomized trials; In observational studies the assumptions may hold to varying degrees. 

For instance, A2, which is also sometimes referred to as the overlap condition can be directly assessed. However, by comparison A3 is untestable and therefore indirect techniques are needed to determine the degree to which it is satisfied most commonly via sensitivity analyses \citep{rosenbaum1982assessing}. These assumptions are necessary for the formal results in the transformed response variable framework to hold.

Beyond these, we make one additional assumption that allows us to simplify the statistical the model we specify in this paper: Stable Unit Treatment Value Assumption (SUTVA) --- This condition assumes no interference between observations, and that there are no multiple versions of the treatment (A4). In its absence, we would need to define a different potential outcome for the unit of observation not just for each treatment received by that unit but for each combination of treatments received by every other observation in the experiment. Relaxing these assumptions will be discussed in Section \ref{futurework} as an avenue that our future work will aim to explore.


The causal estimands considered are the conditional average treatment effect (CATE), that we denote as $\tau^{CATE}$ and the average treatment effect (ATE) that we denote as $\tau^{ATE}$. $\tau^{CATE}$ is the primary estimate of interest in modeling heterogeneous treatment effects and is defined as,
\begin{equation}
\label{eq:cate}
\tau^{CATE} = \mathbb{E}_{Y}[Y(1) - Y(0) \mid X = x],
\end{equation}
the ATE can be derived by integrating over the the joint distribution of the covariates
\begin{equation}
\label{eq:ate}
\tau^{ATE} = \mathbb{E}_{X}\Big[ \mathbb{E}_{Y}[Y(1) -Y(0) \mid  X =x ] \Big]  =\mathbb{E}_{X}[\tau^{CATE}].
\end{equation}

The idea behind the transformed response variable apporach is to define a variable $Y_i^{*}$ for which the conditional expectation with respect to the response recovers the CATE under A3 (see Appendix \ref{app:A} for a proof of this result). A transformation that satisfies the above condition is,

\begin{equation}
\label{eq:trv}
Y^{*}_{i}  = f(W_{i}, Y_{i}, e_{i})= \frac{W_{i} - e_{i}}{e_{i}(1-e_{i})} Y_{i}.
\end{equation}

The transformation requires knowledge of the  probability of receiving the treatment. We assume that the treatment assignment probability depends on the observed covariate levels, or  $e_{i} = e_{i}(X = x_{i})$ is a propensity score. A trivial example is when the propensity score is a fixed covariate independent value, $e_{i} = e$. This is not an example commonly seen in real observational causal inference problems and is as such not considered as a part of the model presented here, albeit  \citep{athey2015machine, Athey:2015:MLC:2783258.2785466, athey2016recursive} consider it as a means of model validation.

\subsection{Strengths and Weaknesses of Past Work in TRV Modeling}\label{LMS}
TRV modeling offers three main advantages when used for estimating treatment effects as demonstrated in prior studies. Foremost amongst these is that the TRV can easily be modeled with any  supervised learning method. For instance, regression trees and random forests have been used \citep{athey2015machine,Athey:2015:MLC:2783258.2785466,wager2017estimation} as has boosting  \citep{powers2017some}. This is not an exhaustive list, and there are a myriad of other methods that can be used in conjunction with the TRV to estimate heterogeneous treatment effects. Furthermore, relative to conditional mean regression, this method does not ignore the propensity score which explicitly enters the estimation via the transformation. Finally, based on the modeling approach used, we can address treatment effect heterogeneity flexibly and therefore avoid issues arising from model misspecification since it is likely that there are complex relationships between the covariates and heterogeneity of the treatment effects. Despite their usefulness, the TRVs have some key weaknesses. 

First, as mentioned in \cite{athey2015machine} and \cite{powers2017some} using TRVs as CATE estimators results in high variance estimates of the causal estimands. By construction the treatment assignment probability and the assignment itself only enter the model implicitly via the transformation and are therefore only accounted for indirectly. In addition, the treatment assignment probability only appears in the denominator, and if this is close to zero or one, the variance can spike. Similar difficulties have been seen in IPW \citep{hirano2003efficient} estimators, that like this transformation grant more weight to tail (read: unlikely) observations. Combining supervised learning techniques with inverse-probability weighting, gives rise to double-robust estimators, which in spirit is also similar to our modeling of the transformed outcome.
\cite{ding2017causal} summarize that the instability of the estimator due to extreme treatment assignment probabilities is even worse in this case than in inverse-probability weighting, since there are potentially two sources of model misspecification. While we can address concerns of model misspecification using flexible machine-learning models, this flexibility is a double-edged sword. When the model generates predictions that are inherently high variance such as those of regression trees, this means that the method suffers in terms of efficiency and the quality of inference is degraded.

Second, uncertainty quantification using methods built atop inverse-probability weighting in general and transformed outcomes in specific is difficult. As discussed at length earlier, there are theoretical concerns due to the the impact of extreme weights which is a limitation of the transformation. There are also practical concerns with uncertainty quantification under specific models for the TRV as it relates to generating intervals. For single regression trees as well as the other ensemble learning methods which have been used for TRV modeling, intervals have been generated using the bootstrap. Prior work \citep{wager2014confidence} has suggested that in certain applications the Monte Carlo error can dominate the uncertainty quantification produced. In conjunction with the high variance inherent to the aforementioned approaches, we might be unable to gather useful insights. If treatment effects are small (near zero), the conflation of the Monte Carlo noise with the underlying sampling noise may lead us to overstate the variance and therefore lower the power of our analysis. In addition note that when the sample size is small, \cite{powers2017some} demonstrate that the variance of the TRV is small as well --  it increases with increasing sample size. Hence, in situations where bootstrapping is likely to do well for the uncertainty in the model i.e. in large samples, the high variance of the TRV is even more so an issue. 

Based on these limitations, we propose the Gaussian process mixture model in Section \ref{model}. Our proposed model attempts to overcome the aforementioned limitations by leveraging the mixture distribution implied by the transformation. In addition, we still aim to model the TRV flexibly and capture the complexity of treatment effect heterogeneity. We achieve gains in the quality of inference by constructing a likelihood that reflects the error structure imposed by the TRV under some basic assumptions that earlier work with this technique has ignored. The details of these findings will be discussed in greater depth in Section \ref{data} where these approaches are applied to real and simulated data. 

\section{The Gaussian Process Mixture Model}\label{model}

We specify a non-parametric Bayesian model based on a mixture of Gaussian processes to model heterogeneous treatment effects. Our model is based on the transformed response variable framework. It is motivated by three objectives: (1) to explicitly model the distribution implied by the transformed outcome with the goal of reducing the variance of the TRV generated estimates that have hitherto been produced using non-probabilistic models, (2) model the two treatment groups jointly so we can borrow strength and therefore improve inference even relative to non-TRV based methods for estimating treatment effects, and (3) making more complete use of the data by jointly modeling the transformed response as well as the treatment assignment probabilities in a one step model. The \emph{feedback} between the two stages in joint modeling can improve the point estimation of treatment effects and the propensity scores \citep{zigler2013model}. Throughout this section we assume A1-A4 are satisfied.

\subsection{Model Specification}
\label{understanding}

A natural starting point is to consider two non-parametric regression functions for the response under treatment and control, respectively
\begin{eqnarray*}
Y_{i}(1) &=&   f_{1}(x_{i}) + \varepsilon_{i}(1), \quad   \epsilon_{i}(1)  \stackrel{\mathrm{iid}}{\sim} \mathrm{N}(0, \sigma^{2}), \\
Y_{i}(0) &=&  f_{0}(x_{i}) + \varepsilon_{i}(0), \quad   \epsilon_{i}(0)  \stackrel{\mathrm{iid}}{\sim} \mathrm{N}(0, \sigma^{2}).
\end{eqnarray*}

In expectation, the difference of these two non-parametric functions is the conditional average treatment effect. Substituting these non-parametric regression functions under the treatment and control cases in the definition of the TRV in \eqref{eq:trv} yields the following mixture model,

\begin{equation}
\label{eq:newmodel}
Y_{i}^{*} = g(x_{i}) + \varepsilon_{i}^{*},
\end{equation}
$$
\varepsilon_{i}^{*} \sim e_{i} \mathrm{N}\bigg((1-e_{i})  h(x_{i}), \frac{1}{e_{i}^{2}}\sigma^{2}\bigg) + (1-e_{i})\mathrm{N}\bigg(-e_{i} h(x_{i}), \frac{1}{(1-e_{i})^{2}}\sigma^{2}\bigg).
$$

where $g(\cdot)$ is interpreted as the conditional average treatment effect,

$$g(x_{i}) = f_{1}(x_{i}) - f_{0}(x_{i}).$$

while the function $h(\cdot)$, helps expresses the multi-modal nature of the error distribution that is implied by the transformation,

$$h(x_{i}) = \frac{f_{1}(x_{i})}{e_{i}} + \frac{f_{0}(x_{i})}{1-e_{i}}.$$ 
 
A detailed derivation of this model is given in Appendix \ref{app:B}.  

The argument for specifying the TRV mixture model rather than individual models for the treatment and control is that the conditionals $Y_i  \mid X_i, W_i =1$ and
$Y_i  \mid X_i, W_i =0$  may not be perfectly estimable. Past work has indicated that ignoring shared information between the treated and untreated groups is a potential source of bias in the treatment effect estimation  \citep{powers2017some}. Under the Bayesian paradigm, methods that place individual vague priors on the aforementioned conditionals make it challenging to control the degree of heterogeneity the model adapts to since the implied priors on their differences is potentially extremely vague \citep{hahn2017bayesian}.

Our model formulation can be considered under two specifications -- when the treatment assignment probabilities are known and when they need to be inferred from the data. The details of each specification are given in Sections \ref{simspec} and \ref{compspec1} for the two cases respectively.

\subsubsection{Model specification with known assignment probabilities}
\label{simspec}
We will place Gaussian process priors on both $g$ and $h$ and will specify an inverse gamma prior on $\sigma^2$ to leverage conjugacy. Therefore, for the case where the treatment probabilities are known we specify the following model,

\begin{equation}
\label{basicmodel}
\begin{aligned}
Y_{i}^{*} &= g(x_{i}) + \varepsilon_{i}^{*},\\
\varepsilon_{i}^{*} \sim e_{i} \mathrm{N}\bigg((1-e_{i})  h(x_{i}), \frac{1}{e_{i}^{2}}\sigma^{2}\bigg) &+ (1-e_{i})\mathrm{N}\bigg(-e_{i} h(x_{i}), \frac{1}{(1-e_{i})^{2}}\sigma^{2}\bigg),\\
g & \sim\mathrm{GP}(0, \kappa_{g}), \\
h & \sim\mathrm{GP}(0, \kappa_{h}), \\
\sigma^2 & \sim \mathrm{IG}(a,b).
\end{aligned}
\end{equation}

Here $\mathrm{IG}(a, b)$ is the inverse gamma distribution with hyper-parameters $a$ and $b$ and $\mathrm{GP}(\textbf{0}, \boldsymbol{\kappa})$ denotes the Gaussian process priors on the function $g$ and $h$. Both priors are zero mean and have covariance kernels specified (1) a non-stationary linear kernel $\kappa_{g}(u, v) = s^{2}_{0} + \sum_{i=1}^{p}s^{2}_{i}(u_{i}-c_{i})(v_{i}-c_{i})$, with hyper-parameters $s^{2}_{0}, \ldots s^{2}_{p}$ on $g$ and (2) a square exponential, $\kappa_{h}(u, v) = s_{h}^{2}\exp\{- \tau^2 \| u-v\|^2\} $ with hyper-parameters $\tau, s^{2}$ on $h$. These kernels rely on the notion of similarity between data points -- if the inputs are closer together than the target values of the response, in this case the TRV are also likely to be close together. Under the Gaussian process prior, the kernel functions described above formally define what is near or similar.

The hyper-parameters $s^{2}_{0}, \ldots s^{2}_{p}$  can be interpreted in the context of linear regression with $\{\mathrm{Normal}\sim(0, s^{2}_{j})\}_{j=0}^{p}$ priors on the $p+1$ regression coefficients including the intercept. The offset $\{c_{i}\}_{i=1}^{p}$ determines the $x$ coordinate of the point that all the lines in the posterior is meant to go through. This provides some insight into how these can be set for applied modeling problems. In cases where there is a large number of covariates, many of which are thought to share information, the prior variance for those dimensions can be made small, with a higher degree of mass concentrated near zero to induce more shrinkage. In contrast, where there is a small number of important covariates the prior variance can be set to make the prior more diffuse. The offset can be set to the average of each covariates observed value. This is a general overview of the strategy that we have employed.

\subsubsection{Model specification with unknown assignment probabilities}
\label{compspec1}
Computing the TRV requires knowledge of the treatment assignment probabilities $\{e_i\}_{i=1}^{n}$. In the case where these are unknown we consider them as latent variables and add extra levels to the hierarchical model specified in \eqref{basicmodel} to model the treatment assignment probabilities. We model the assignment probabilities individually so for notational ease, later in this paper we use $\pmb{e} = \{e_i\}_{i=1}^{n}$. Our specification, \emph{apriori}, assumes that the assignment mechanism and the outcome model are independent.

\paragraph{Modeling the Propensity Score} 

In order to learn the treatment assignment probabilities, we specify a probit regression model that is layered onto the model defined in \eqref{basicmodel}.
\begin{equation}
\label{indassign}
\begin{aligned}
W_{i} &\sim\mathrm{Ber}(e_{i}),\\
e_{i} &= \Phi(X_{i} \boldsymbol{\beta}), \\
\boldsymbol{\beta} & \sim\mathrm{N}_{p+1}(0, \Psi_{p+1 \times p + 1}).
\end{aligned}
\end{equation}
Where $\Phi$ denotes the standard Normal cumulative distribution function. In this paper we will only consider the above Gaussian prior on $\boldsymbol \beta$ with prior covariance $\Psi$. However, additional complexity can be added by allowing the coefficient vector $\boldsymbol \beta$ to vary via a hierarchical prior structure as may be motivated by more complex multi-stage clustered data. 

\subsection{Posterior Sampling with Known Assignment Probabilities}
\label{inference}

Inference for the model specified in Section \ref{simspec} involves sampling from a posterior distribution via straightforward Gibbs-sampling.

We define $\mathbf{g} = (g(x_{1}), \ldots, g(x_{n}))$ and $\mathbf{h} = (h(x_{1}), \ldots, h(x_{n}))$ as the values of the two regression functions on the training data.
We  denote the TRV as $\textbf{Y}^{*} = (Y_{1}^{*}, \ldots, Y_{n}^{*})$ .  In this case the target joint posterior distribution is 
\begin{equation}
\label{eq:jd1}
\pi(\mathbf{g}, \mathbf{h}, \sigma^{2} \mid \mathcal{D}).
\end{equation} 

Due to prior conjugacy the conditional distributions: $\pi(\mathbf{g} \mid \mathbf{h}, \sigma^{2}, \mathcal{D})$, $\pi(\mathbf{h} \mid \mathbf{g}, \sigma^{2}, \mathcal{D})$   and $\pi(\sigma^{2} \mid \mathbf{h}, \mathbf{g}, \mathcal{D})$ all have simple forms that we can easily sample from. We first state some matrices and vectors that will enter our calculations: $\mathbf{D}$ is an $n\times n$ diagonal matrix with entries $\mathbf{D}_{ii} = \bigg( \frac{W_i }{e_{i}^{2}} \sigma^2 + \frac{1-W_i}{(1-e_{i})^{2}}\sigma^{2}\bigg)$, $\boldsymbol{\Lambda}$ is also an  $n\times n$ diagonal matrix with entries $\boldsymbol{\Lambda} _{ii} = \bigg(W_{i}(1-e_{i}) + (1-W_{i})(-e_{i})\bigg)$, $\mathbf{K}$ is also an  $n\times n$ diagonal matrix with entries $\mathbf{K}_{ii} = \sigma^{2}\mathbf{D}_{ii}$, and $\mathbf{m} = \Lambda \textbf{H}$. We also denote the covariance matrix
$\pmb{\kappa}_{g}$ with the $ij$-th entry as taking the value $\kappa_g(x_i,x_j)$ and similarly $\pmb{\kappa}_{h}$ is a matrix with the $ij$-th entry taking the value $\kappa_h(x_i,x_j)$. We now state the conditional distributions that will enter our Gibbs sampler,
\begin{equation}
\label{eq:fcg} 
\begin{aligned}
\pi(\mathbf{g} \mid \mathbf{h}, \sigma^{2}, \mathcal{D})  &\sim \mathrm{N}\big((\pmb{\kappa}_{g}^{-1}+\mathbf{D}^{-1})^{-1}(\mathbf{D}^{-1}\textbf{Y}^{*} - \mathbf{m}\}, \{\pmb{\kappa}_{g}^{-1}+\mathbf{D}^{-1})^{-1}\big), \\
\pi(\mathbf{h} \mid \mathbf{g}, \sigma^{2}, \mathcal{D}) &\sim \mathrm{N}\bigg((\pmb{\kappa}_{h}^{-1} + \boldsymbol{\Lambda}^{T} \mathbf{D}^{-1} \boldsymbol{\Lambda})^{-1} \boldsymbol{\Lambda}^{T} \mathbf{D}^{-1} (\textbf{Y}^{*}-\mathbf{g}), (\pmb{\kappa}_{h}^{-1} + \boldsymbol{\Lambda}^{T} \mathbf{D}^{-1} \boldsymbol{\Lambda})^{-1} \bigg),  \\ 
\pi(\sigma^{2} \mid \mathbf{h}, \mathbf{g}, \mathcal{D}) & \sim  \mathrm{IG} \bigg(a+\frac{n}{2}, b + \frac{(\textbf{Y}^{*}-\mathbf{g} - \mathbf{m})^{T}\mathbf{K}^{-1}(\textbf{Y}^{*}-\mathbf{g}- \mathbf{m})}{2}\bigg).
\end{aligned}
\end{equation}
The Gibbs steps that would be used to sample from these full conditional distributions are given appendix \ref{app:D}.

\subsection{Posterior Sampling with Unknown Assignment Probabilities}
\label{complexinference}

There are two additional problems with respect to inference when the assignment probabilities are unknown: one needs to estimate the assignment probabilities $\pmb{e}$ and use these to compute the TRV $\textbf{Y}^{*}$. The following target posterior distribution corresponds to the model when the treatment probabilities are modeled as specified by the probit augmentation to the model in \eqref{indassign}.

\begin{eqnarray}
&\pi(\mathbf{g}, \mathbf{h}, \textbf{Y}^{*}, \sigma^{2},\boldsymbol{e}, \boldsymbol{\beta} \mid \mathcal{D}). \label{post3}
\end{eqnarray}

In this setting the joint posterior is more complicated than equation \eqref{eq:jd1} and is harder to sample from since it cannot be completely decomposed into Gibbs steps. Generating samples requires incorporating the full conditional distributions from the previous section, along with additional steps to sample the treatment assignment probability by using a Metropolis-within-Gibbs step and constructing the transformed outcome. 

The Metropolis-Hastings step consists of specifying a proposal distribution $q(\boldsymbol{\beta})$, and given a candidate value $\pmb{\beta}^* \sim q(\pmb{\beta})$ is accepted with acceptance probability,
\begin{equation}
\label{eq:mhratio}
\alpha = \min\left(1, \frac{\pi(\mathbf{g}, \mathbf{h}, \textbf{Y}^{*},\sigma^{2}, \boldsymbol{\beta}^{*}, \pmb{e} \mid \mathcal{D}) \, q(\pmb{\beta})}{\pi(\mathbf{g}, \mathbf{h}, \textbf{Y}^{*}, \sigma^{2}, \boldsymbol{\beta}, \pmb{e} \mid \mathcal{D}) \, q(\pmb{\beta}^{*})}\right).
\end{equation}
where the posterior for evaluation is given in \eqref{post3}. We have used a symmetric random walk proposals\footnote{We generate proposals as $\pmb{\beta}^{*}\sim\mathrm{N}(\mu = \pmb{\beta}^{j-1}, \sigma^{2})$ i.e. from a Gaussian distribution that is centered at the last accepted value. The variance controls the step size of the proposals and needs to be tuned for the application.} in order to reduce the overall computational burden. Once we have sampled the coefficients for the probit model, we can deterministically compute the treatment assignment probability and the TRV. The complete algorithm for this sampling scheme is detailed in appendix \ref{app:D}.

\paragraph{Joint Bayesian modeling and the feedback problem:} The joint Bayesian model specified in this paper for learning the assignment mechanism $\pmb{e}$ and the transformed outcome $\textbf{Y}^{*}$ leads to a feedback problem of the type described in \cite{zigler2013model}. The treatment assignment probability $\pmb{e}$ appears in the joint posterior distribution both as a part of the transformed outcome model through \eqref{eq:fcg} as well as its own model in \eqref{indassign}. Therefore its posterior samples involve information from both. In the specific context of the assignment model, this means that the posterior samples of parameters in learning $\pmb{e}$ are informed by information from the outcome stage. 

Under the classical method of using $\pmb{e}$ as a dimension reduced covariate representation in the outcome stage model (an analog to our transformed outcome), \citep{zigler2013model} demonstrate that the estimation of causal effects is poor. There is a possibility of considerable bias due to the distortion of the causal effects. Furthermore, the usefulness of the propensity score adjustment as a replacement for the covariates is also compromised. 

However, this is not the concern in the modeling scheme proposed in this paper. \cite{zigler2013model} show that the nature of the feedback between the two stages is altered when the outcome stage model is augmented with adjustment for the individual covariates and that this method can recover causal effects akin to when a classical two stage procedure is used. Our approach via the kernels of the Gaussian processes provides individual covariate adjustment therefore alleviating concerns created by the feedback. Therefore we reap the benefits of the joint estimation, but by means of suitably elicited priors, and individually controlled covariates, we bypass the concerns of feedback. In fact, by making more complete use of the data, we are arguably able to improve the overall quality of estimation.

\section{Results on Simulated and Real Data}
\label{data}

In this section we validate our Gaussian process based TRV model on simulated and real data. We use the simulations to show that our approach outperforms other techniques (both TRV as well as conditional mean regression type methods). This holds true both when the treatment assignment probabilities are known or need to be inferred from the data. We also observe on the simulated data that our model does in fact recover the causal effects in the TRV framework in the presence of feedback as theorized earlier. Our assertion is based on comparisons of mean squared error, bias and point-wise coverage of the uncertainty intervals generated by the model. 

The real data analyzed here comes from a study of the causal effects of debit card ownership on household spending in Italy \citep{mercatanti2014debit} -- we will refer to these data as the SHIW data. In the analysis of the SHIW data we jointly infer treatment effects as well as the treatment assignment probability for each individual, as these are not observed.

The most interesting aspect of our analysis of the SHIW data is that we are able to identify heterogeneity in the treatment effects. We find that the impact of debit card usage on aggregate household spending is found to vary based on income and this variability is highest at the lowest levels of income -- a notion that is validated under behavioral economic theory which further lends credibility to our proposed model.

\subsection{Estimands Used and Modelling Approaches Compared}

In this section we state the estimands that we will use for comparing our method to other non-parametric methods. We will also state in detail how we compute the relevant estimand for both our method and the other techniques considered. The analysis is focused on the estimation of the CATE.

\paragraph{Gaussian process mixture model:} We first specify the procedure we use to estimate the CATE for our model. The model is trained on data  $(x_1,...,x_n)$  and the values of the two functions are
\begin{eqnarray*}
{\mathbf{g}} = (g(x_{1}), \ldots, g(x_{n})), \\
{\mathbf{h}} = (h(x_{1}), \ldots, h(x_{n})).
\end{eqnarray*}
We will use the function values to evaluate the accuracy of our estimators.

Depending on whether the treatment assignment probabilities are observed or not we obtain posterior samples $\left(\mathbf{g}^{(j)},\mathbf{h}^{(j)}\right)_{j=1}^K$ 
or $\left(\mathbf{g}^{(j)},\mathbf{h}^{(j)}, \boldsymbol{e}^{(j)}\right)_{j=1}^K$, respectively, using which we can compute posterior samples for the conditional average treatment effect at each location $x_i$, $i=1,...,n$ as
\begin{equation*}
{\tau_{i}^{CATE}}^{(j)}(x_i) = g^{(j)}(x_i).
\end{equation*}
Given the posterior samples we can compute a posterior mean as a point estimate, 
$\widehat{\tau_{i}^{CATE}}$ along with its corresponding credible intervals. Where applicable, marginalizing over the values $x_{i}$ allows us to compute posterior estimates of the average treatment effect $\widehat{\tau_{i}^{ATE}}$.

Based on the quantities that we have specified above, the mean squared error, bias and coverage used for model validation are specified as follows,

\begin{equation*}
\mathrm{Mean \>\>\> Squared \>\>\> Error}  = \frac{1}{n}\sum_{i=1}^{n}(\tau_{i}^{CATE}-\widehat{\tau_{i}^{CATE}})^{2},
\end{equation*}

\begin{equation}
\mathrm{Bias} = \frac{1}{n}\sum_{i=1}^{n}(\tau_{i}^{CATE}-\widehat{\tau_{i}^{CATE}}),
\end{equation}

\begin{equation*}
\mathrm{Coverage} =  \frac{1}{n}\sum_{i=1}^{n} \textbf{1}(\tau_{i}^{CATE}\in[\tau_{i}^{CATE, \>\>\> lwr}, \tau_{i}^{CATE, \>\>\> upr}]).
\end{equation*}

\paragraph{Summary of alternative methods used:} We will compare our proposed Gaussian process mixture model approach to other regression based methods for estimating treatment effects. We have considered random forests and single regression trees for treatment effect estimation via TRV modeling as well as \emph{fit based trees}, \emph{causal trees} \citep{athey2016recursive}, and BART\citep{hahn2017bayesian, hill2011bayesian} as non-TRV alternatives \footnote{We use the implementations of these methods in the \texttt{R} packages \texttt{causalTree} \citep{causal2016tree}, \texttt{rpart}\citep{rpart2002}, \texttt{randomForest}\citep{rf2018} and \texttt{BART}\citep{bart2018}}.

None of the aforementioned methods have an obvious framework for learning the treatment assignment probabilities internally. This a crucial step in computing the CATE and ATE both via TRV and non-TRV based estimation techniques. In the case of the regression trees and random forests for TRV modeling, the TRV needs to be computed from the learnt propensity score first before any modeling can commence. The BART model uses the propensity score as an additional covariate, while causal and fit based trees use the propensity score as a weighting mechanism.

Therefore, we will use a two-step procedure where we first use the data to infer the treatment assignment probabilities and then given these estimates, apply the aforementioned regression methods to estimate the treatment effect. The treatment assignment probability vector  $\pmb{e}$  is estimated via logistic regression \citep{rubin1996matching}, a standard approach for estimating propensity scores in the causal inference literature.

\subsection{Results on Simulated Data}

The objective of the simulation studies presented in this section is to compare the performance of the Gaussian process mixture model to, BART, causal trees, fit based trees, the random forest and single regression tree models. We consider two criteria in our comparison. The first criteria is a comparison of the accuracy of the CATE, in terms of mean squared error and bias. The second criterion involves assessing how well the methods quantify uncertainty by considering the coverage of the intervals produced by all the models.

\subsubsection{Simulated Data Model}
\label{simstudy}

In order to evaluate the proposed model as well as the other aforementioned approaches, we consider two simulation settings -- one high dimensional case (with 40 covariates) and one low dimensional case (with 5 covariates) each with its own covariate level heterogeneity and a sample size of $n = 250$. For the remainder of this analysis, the high dimensional case is referred to as Case A, and the low dimensional case is referred to as Case B. By design neither of these simulation cases has a meaningful average treatment effect. We start with a detailed description of Case A.

In this framework, covariates $X_{1}, \ldots X_{30}$ are independent covariates, $X_{31}, \ldots X_{35}$ depend on pairs of covariates, while $X_{36}, \ldots, X_{40}$ depend on groups of three as follows,

\begin{align*}
X_{k} &\sim \mathrm{Normal}(0, 1); \>\>\> k = 1, \ldots, 15,\\
X_{k} &\sim \mathrm{Uniform}(0, 1); \>\>\> k = 16, \ldots, 30,\\
X_{k} &\sim \mathrm{Bernoulli}(q_{k}); \>\>\> q_{k} = \mathrm{logit}^{-1}(X_{k-30} - X_{k-15}); \>\>\> k = 31,\ldots, 35,\\
X_{k} &\sim \mathrm{Poisson}(\lambda_{k}); \>\>\> \lambda_{k} = 5 + 0.75 X_{k-35}(X_{k-20} + X_{k-5}); \>\>\> k = 36,\ldots, 40.\\
\end{align*}

\begin{figure}[htb]
\centering
\makebox[\textwidth]{
    \includegraphics[scale = 0.33, page = 1]{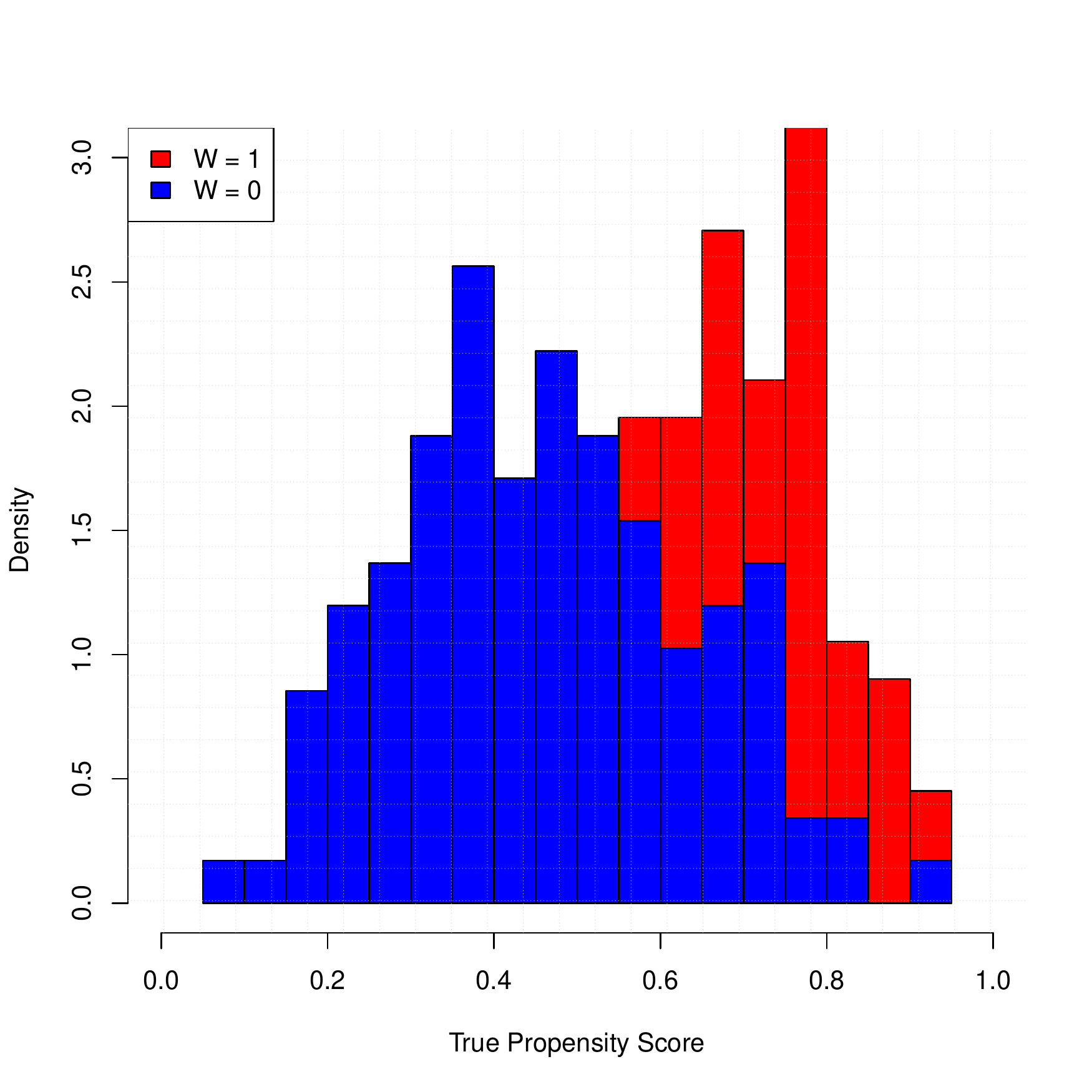}\hspace{-0.2cm} 
    \includegraphics[scale = 0.33, page = 3]{./Results/JMLR/"simPlotsScoreLarge".pdf}
    }
  \\ \vspace{-.25cm}
    \makebox[\textwidth]{ (a) \hspace{2.2in}  (b)
    } 
\caption{Summary plots of Case A (a) Histogram of the true propensity scores for each of the two treatment groups. (b) Treatment effects. The simulation was generated with $n = 250$}
\label{fig:simLarge}
\end{figure}

Next, we simulate the propensity score and the corresponding treatment assignments. This has been done as a simple linear transformation since the focus of the paper is not propensity score modeling but rather CATE modeling. The propensity scores and the treatment effects of interest for Case A are given in figure \ref{fig:simLarge}.

\begin{align*}
p_{i} &= \mathrm{logit}^{-1}(0.3 \sum_{k = 1}^{5}X_{k} -0.5 \sum_{k = 21}^{25}X_{k} -0.0001 \sum_{k = 26}^{35}X_{k} + 0.055 \sum_{k = 36}^{40}X_{k} ),\\
W &\sim \mathrm{Bernoulli}(p_{i}).
\end{align*}
Finally we generate the potential outcomes and the observed outcomes.
\begin{align*}
f(\mathbf{X}) &= \frac{\sum_{k = 16}^{19} X_{k}\exp(X_{k+14})}{1+\sum_{k = 16}^{19}X_{k} \exp(X_{k+14})},\\
Y(0) &= 0.15 \sum_{k=1}^{5}X_{k} + 1.5  \exp(1+1.5 f(\mathbf{X}))+ \epsilon_{i},\\
Y(1) &= \sum_{k = 1}^{5}\{ 2.15 X_{k} + 2.75 X_{k}^{2} + 10 X_{k}^{3}\} + 1.25 \sqrt{0.5 + 1.5\sum_{k = 36}^{40}X_{k}} + \epsilon_{i},\\
Y &= WY(1) + (1-W)Y(0);\>\>\> \epsilon_{i} \stackrel{\mathrm{IID}}{\sim} \mathrm{Normal}(0, 0.0001).
\end{align*}

The lower dimensional case, which we have adapted from the simulation study in \cite{hahn2017bayesian} is presented similarly. We start by simulating the following 5 covariates,
\begin{align*}
X_{k} &\sim \mathrm{Normal}(0, 1); \>\>\> k = 1, \ldots, 3,\\
X_{4} &\sim \mathrm{Bernoulli}(p = 0.25), \\
X_{5} &\sim \mathrm{Binomial}(n = 2, p = 0.5).
\end{align*}

In this scheme, unlike Case A, all the covariates are independent. The propensity score model analogous to the previous case is a linear transformation of the covariates.

\begin{align*}
p_{i} &= \mathrm{logit}^{-1}(0.1X_{1}-0.001X_{2}+.275X_{3}-0.03X_{4}),\\
W &\sim \mathrm{Bernoulli}(p_{i}).
\end{align*}

\begin{figure}[htb]
\centering
\makebox[\textwidth]{
    \includegraphics[scale = 0.33, page = 1]{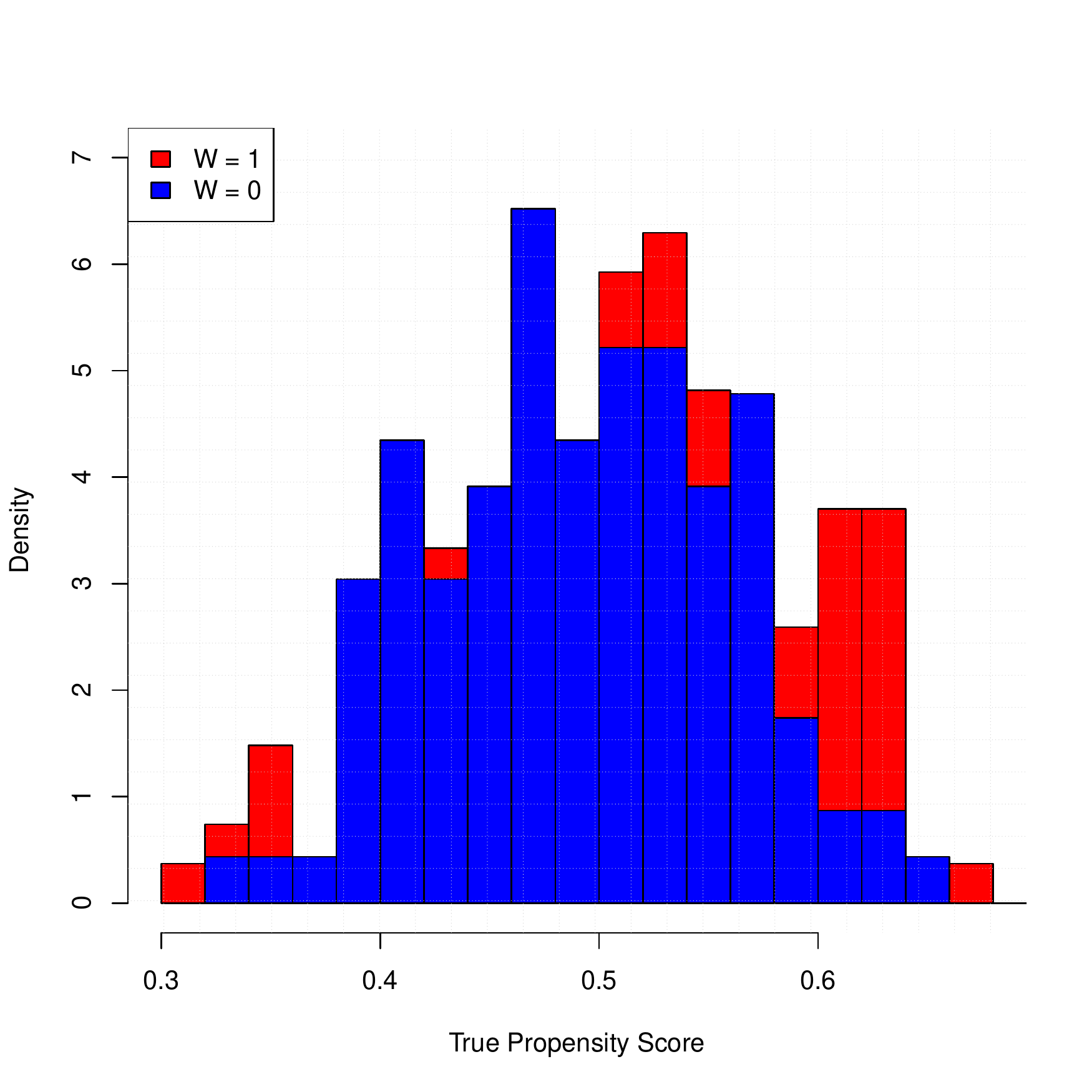}\hspace{-0.2cm} 
    \includegraphics[scale = 0.33, page = 3]{./Results/JMLR/"simPlotsScoreSmall".pdf}
    }
  \\ \vspace{-.25cm}
    \makebox[\textwidth]{ (a) \hspace{2.2in}  (b)
    } 
\caption{Summary plots of Case B (a) Histogram of the true propensity scores for each of the two treatment groups. (b) Treatment effects. The simulation was generated with $n = 250$}
\label{fig:simSmall}
\end{figure}

\noindent Finally we generate the potential outcomes and the observed outcomes. The results of this simulation are presented in figure \ref{fig:simSmall}.

\begin{align*}
f(\mathbf{X}) &= -6 + h(X_{5}) + |X_{3}-1|,\\
h(0) &= 2, \>\>\> h(1) = -1, \>\>\> h(2) = -4,\\
Y(0) &= f(X) - 15 X_{3} + \epsilon_{i},\\
Y(1) &= f(X)+ (1 + 2X_{2}X_{3}) + \epsilon_{i},\\
Y&= WY(1) + (1-W)Y(0); \>\>\> \epsilon_{i} \stackrel{\mathrm{IID}}{\sim} \mathrm{Normal}(0, 0.0001).
\end{align*}

\subsubsection{Comparison of Methods}
\label{ressim}

\begin{figure}[htb]
\centering
\makebox[\textwidth]{
    \includegraphics[scale = 0.35]{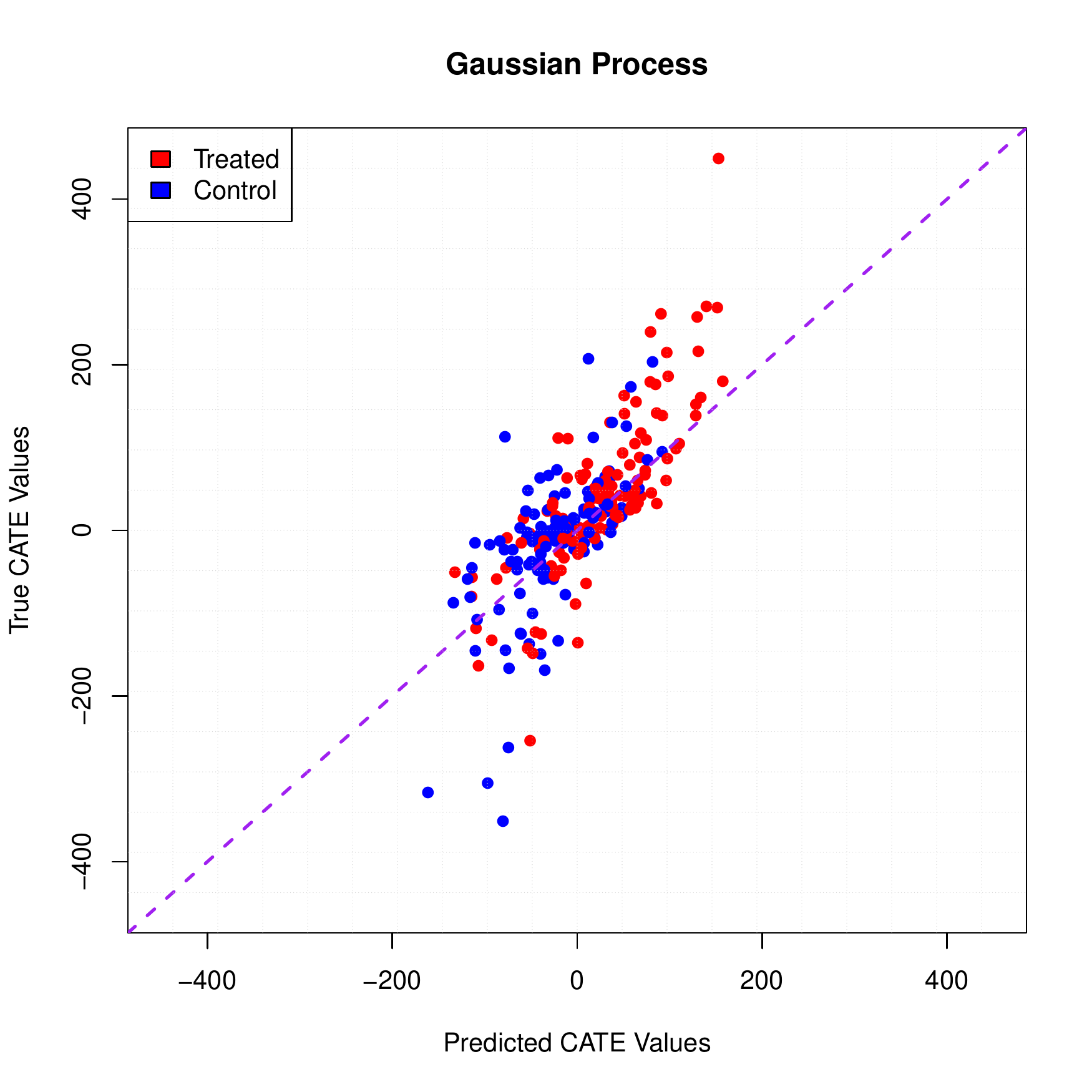}\hspace{-0.4cm}
    \includegraphics[scale = 0.35, page = 1]{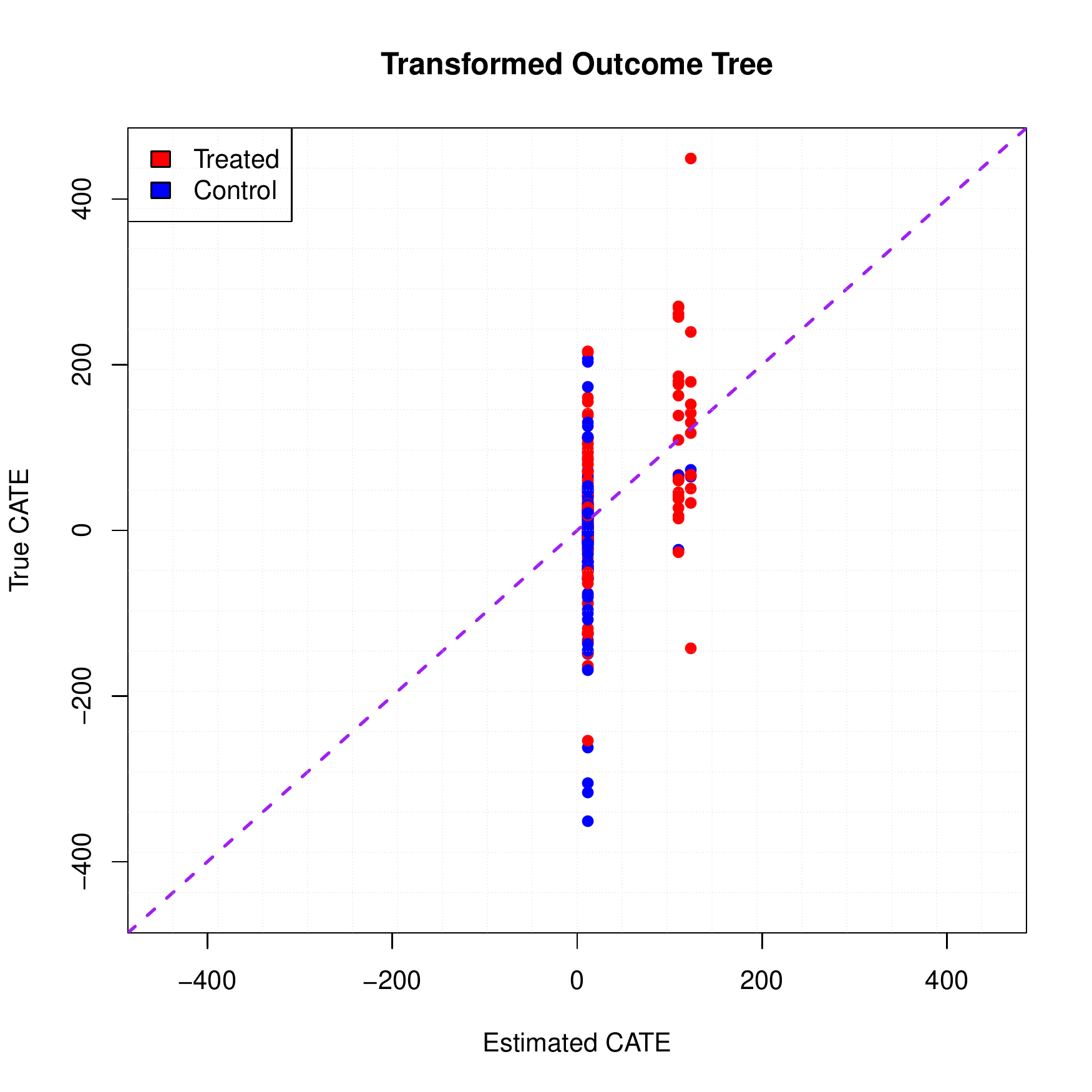}\hspace{-0.4cm}
    \includegraphics[scale = 0.35, page = 1]{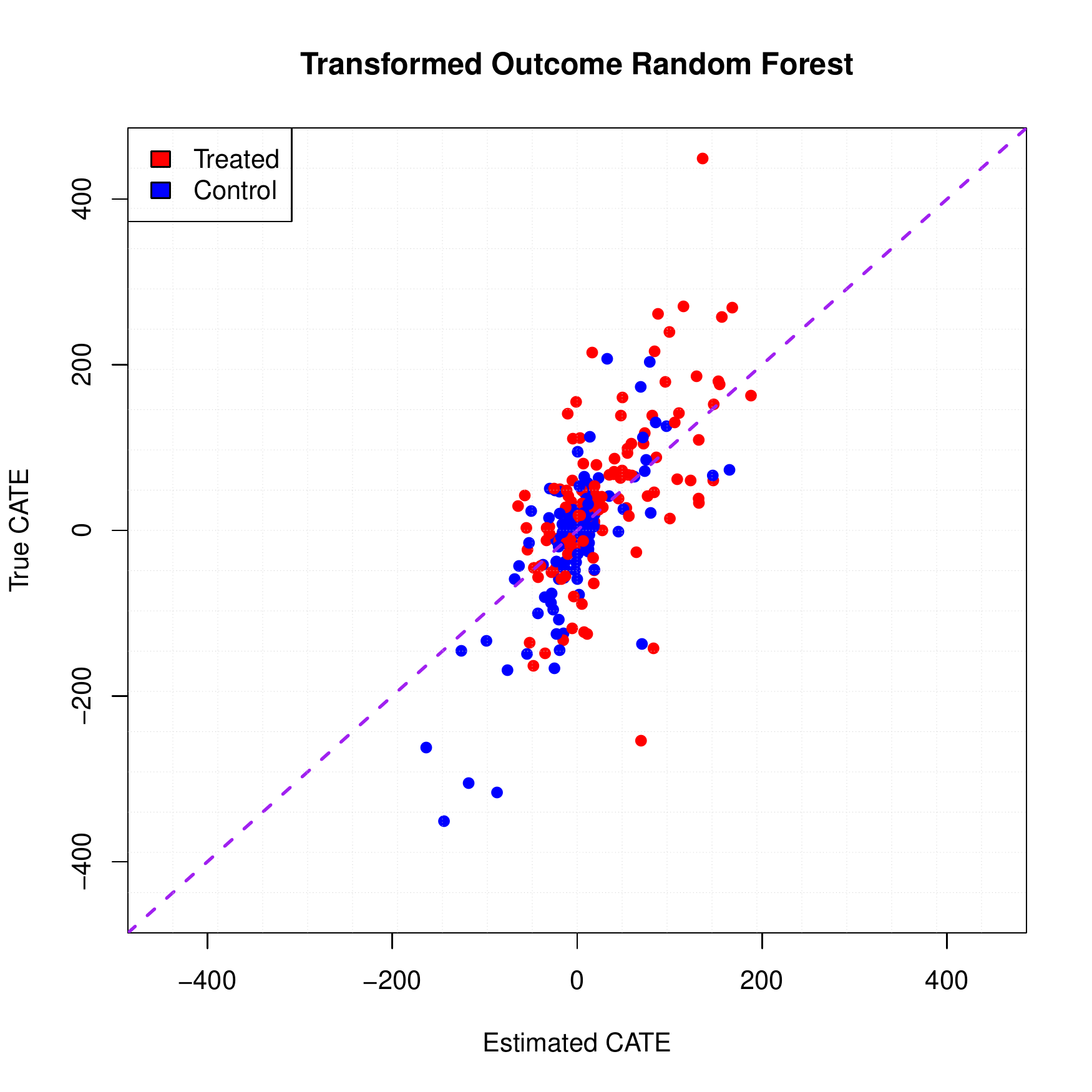}\hspace{-0.4cm}
    } \\ 
    \makebox[\textwidth]{ (a) \hspace{1.8in}  (b) \hspace{1.8in} (c)
    } \\
\makebox[\textwidth]{
    \includegraphics[scale = 0.35, page = 1]{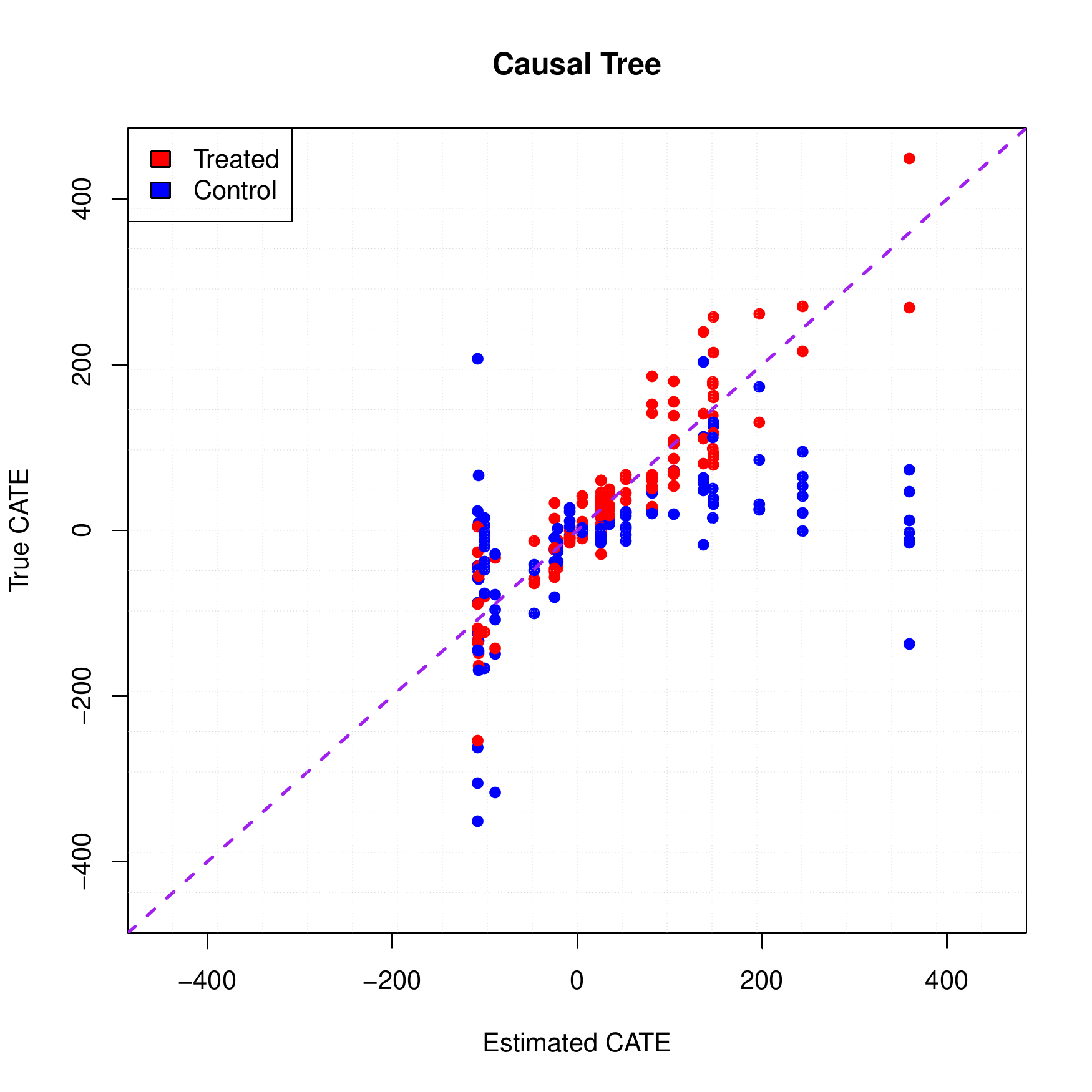}\hspace{-0.4cm}
    \includegraphics[scale = 0.35, page = 1]{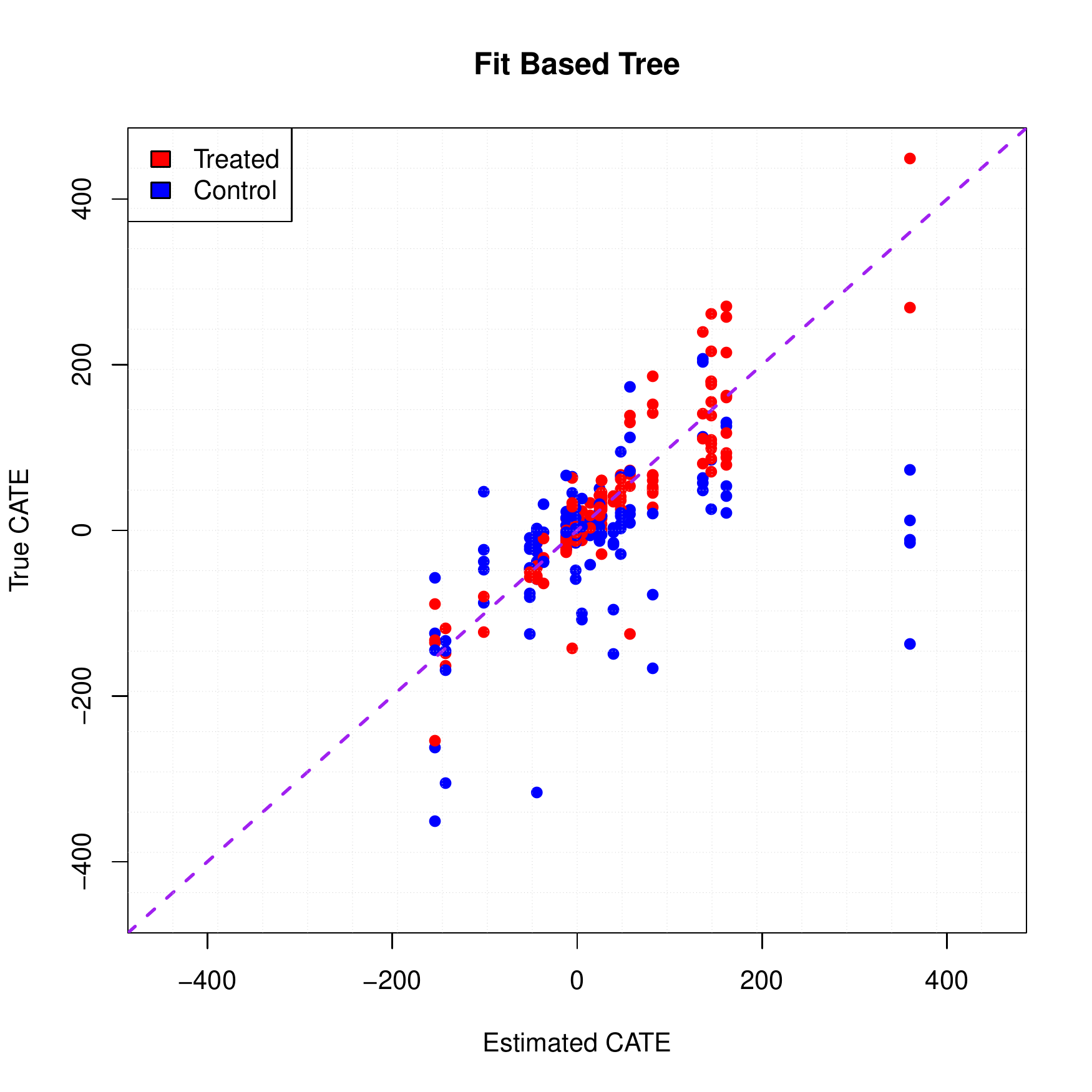}\hspace{-0.4cm}
     \includegraphics[scale = 0.35, page  = 1]{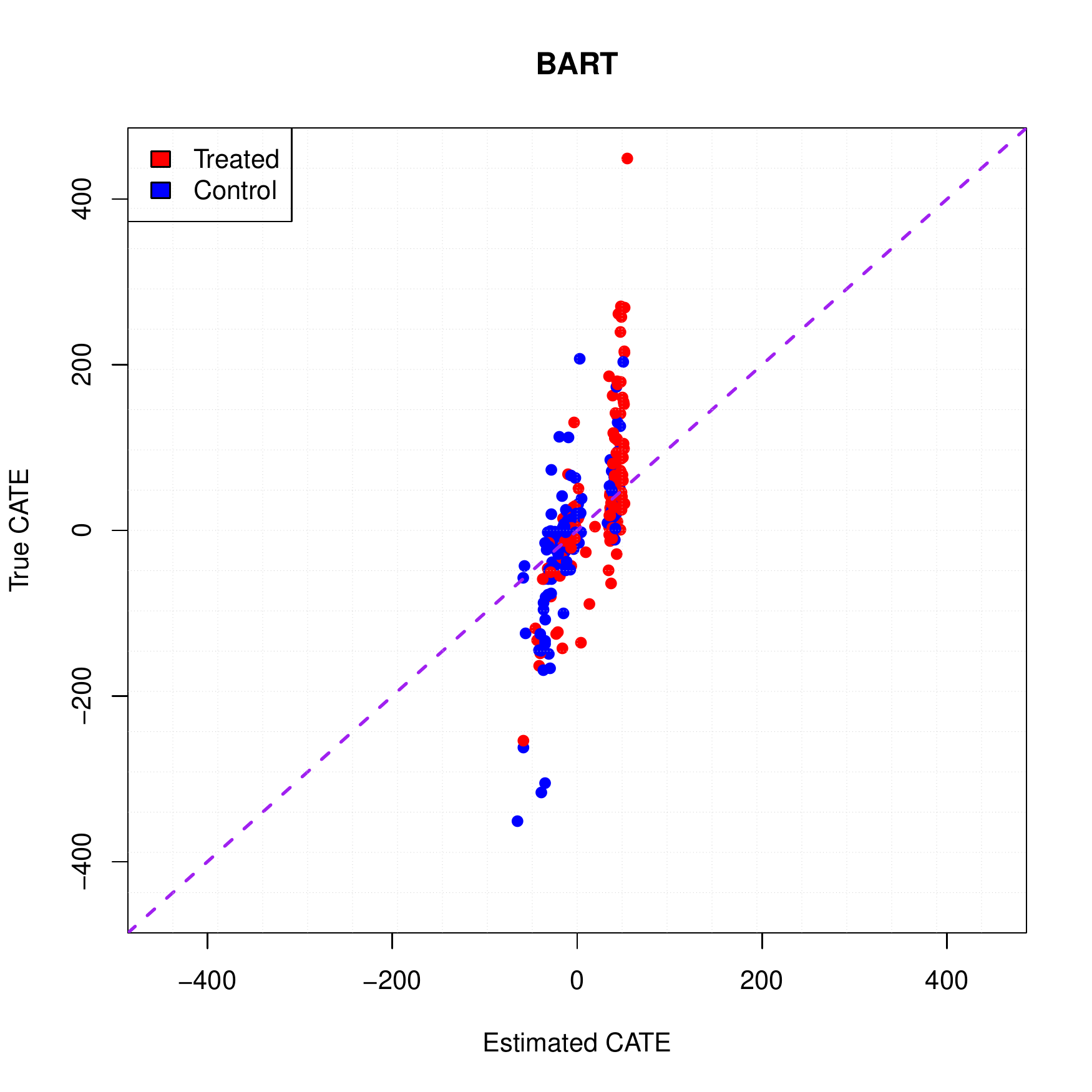}
    
    }\\ 
        \makebox[\textwidth]{ (d) \hspace{1.8in}  (e) \hspace{1.8in} (f) } 
\caption{Comparison of the CATE estimates when the treatment probabilities are known for Case A (a) the GP mixture model (b) the transformed outcome regression tree  (c) the transformed outcome random forest (d) the causal tree (e) fit based tree  (f) BART} 
\label{fig:caseAComparisonKnown}
\end{figure}

\begin{figure}[htb]
\centering
\makebox[\textwidth]{
    \includegraphics[scale = 0.35]{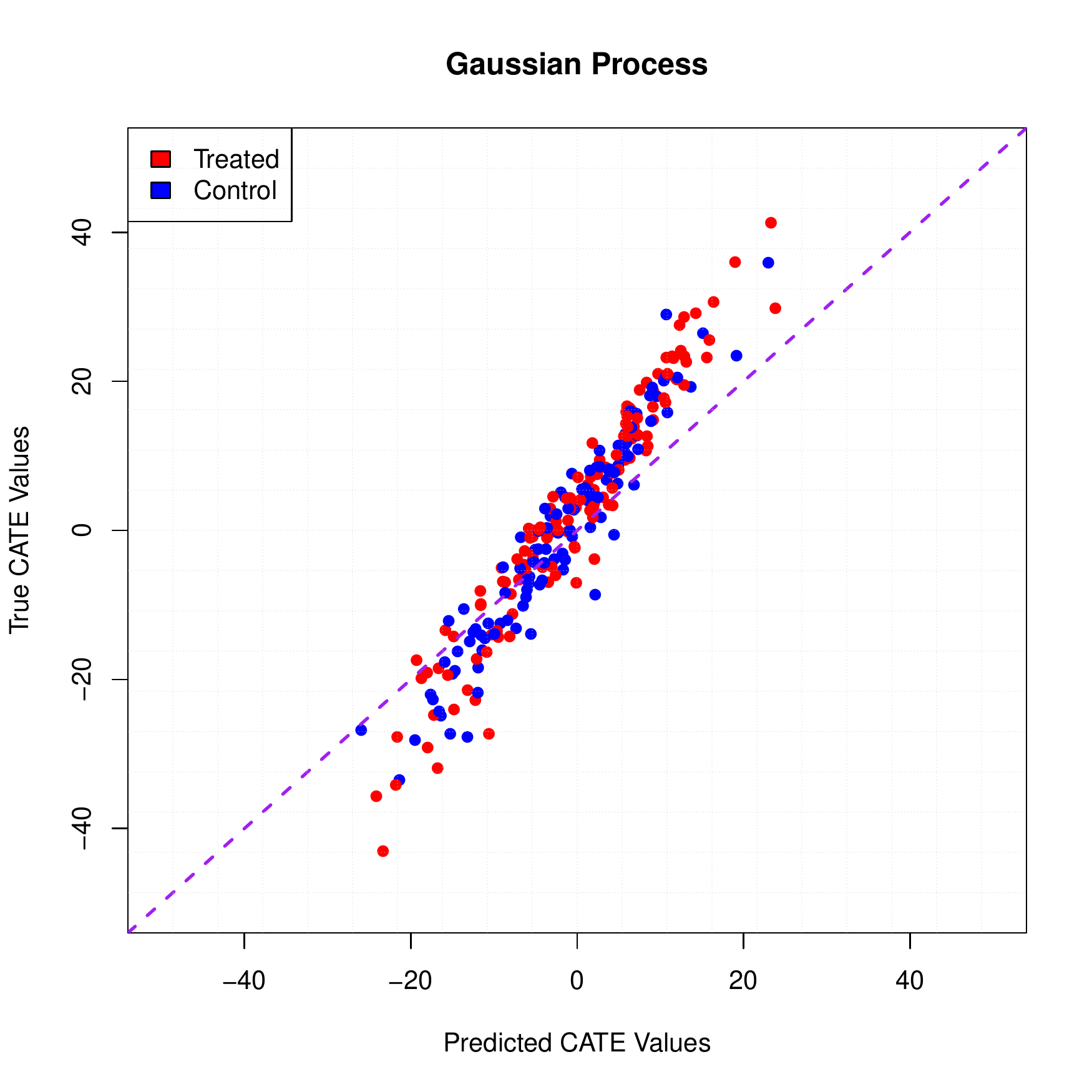}\hspace{-0.4cm}
    \includegraphics[scale = 0.35, page = 2]{"./Results/JMLR/tot_cate_known".pdf}\hspace{-0.4cm}
    \includegraphics[scale = 0.35, page = 2]{"./Results/JMLR/rf_cate_known".pdf}\hspace{-0.4cm}
    } \\ 
    \makebox[\textwidth]{ (a) \hspace{1.8in}  (b) \hspace{1.8in} (c)
    } \\
\makebox[\textwidth]{
    \includegraphics[scale = 0.35, page = 2]{"./Results/JMLR/ct_cate_known".pdf}\hspace{-0.4cm}
    \includegraphics[scale = 0.35, page = 2]{"./Results/JMLR/fit_cate_known".pdf}\hspace{-0.4cm}
     \includegraphics[scale = 0.35, page  = 2]{"./Results/JMLR/bart_cate_known".pdf}
    
    }\\ 
        \makebox[\textwidth]{ (d) \hspace{1.8in}  (e) \hspace{1.8in} (f) } 
\caption{Comparison of the CATE estimates when the treatment probabilities are known for Case B (a) the GP mixture model (b) the transformed outcome regression tree  (c) the transformed outcome random forest (d) the causal tree (e) fit based tree  (f) BART} 
\label{fig:caseBComparisonKnown}
\end{figure}

The first stage of our analysis compares the CATE estimation in instances when the treatment assignment probability is assumed to be known. We focus on the mean squared error, bias and coverage of the CATE under Case A and Case B along with visual analyses of model adaptability to gauge fit quality. For the proposed model the samplers were run for $K = 6, 000$ steps with $1,000$ initial steps burned off. No thinning of the samples generated was needed. Similarly, for the non-Bayesian methods, $K = 5, 000$ replications of the bootstrap were generated. The comparison of point estimates of the CATE under Case A is presented in figure \ref{fig:caseAComparisonKnown} and Case B in figure \ref{fig:caseBComparisonKnown} for the sub-case where the treatment assignment mechanism is known; the corresponding diagnostic measures are presented in tables \ref{tb:caseASummaryKnown} and \ref{tb:caseBSummaryKnown}.

In Case A, both in terms of point estimation, as well as uncertainty quantification, we can conclude that when the treatment assignment is known, the proposed model is the overall winner. As we can see, it adapts well to the heterogeneity of the treatment effects in the data, and is able to recover the effects to a high degree as observed in figure \ref{fig:caseAComparisonKnown}(a). It also has the lowest mean squared error of the models presented and the point-wise coverage of its uncertainty intervals, while low relative to tree based methods, is better than BART (see table \ref{tb:caseASummaryKnown}). Furthermore, the bias of the model is generally lower than causal trees, fit based trees and transformed outcome trees.

It warrants mention that BART only adapts to heterogeneity minimally. We can attribute this to the complexity of regularization in causal inference problems \citep{hahn2017bayesian} from the shrinkage prior as well as poor mixing of the MCMC used for BART in high dimensions \citep{pratola2016efficient}. We see similar behavior from transformed outcome trees, where post-estimation \emph{pruning} can lead to regularization induced bias as well. An elaborate discussion on bias in causal inference applications from regularized models originally designed for prediction is given in \cite{hahn2017bayesian} and \cite{hahn2018regularization}.

In Case B, the model performs well in terms of recovering the high degree of heterogeneity but it suffers in terms of mean square error and bias. The model still adapts well to the heterogeneity inherent in the data, and is able to recover the effects as observed in figure \ref{fig:caseBComparisonKnown}(a), albeit with a higher degree of overall noise. This noisiness translates to high mean squared error and bias, where the other alternative models perform better, with one minor caveat. Due to the piece-wise nature of the tree based models, they do not adapt to the heterogeneity as well as the proposed model and BART do. Furthermore, the model also has the highest degree of point-wise uncertainty interval coverage (see table \ref{tb:caseBSummaryKnown}).

\begin{table}[ht]
\centering
\begin{tabular}{rlrrr}
  \hline
 & Model Type & Mean Square Error & Bias & 95\% CI Coverage \\ 
  \hline
1& Gaussian-Process Mixture & 4191.665 & 13.207 & 0.780 \\ 
  2& Bayesian Additive Regression Tree & 5856.135 & -5.351 & 0.596 \\ 
  3& Transformed Outcome Tree & 7769.077 & 14.374 & 0.876 \\ 
  4& Fit Based Tree & 6154.396 & 15.633 & 0.928 \\ 
  5& Causal Tree & 8390.039 & 21.923 & 0.964 \\ 
  6& Transformed Outcome Random Forest & 4993.576 & 0.317 & 0.932 \\ 
   \hline
\end{tabular}
\caption{Case A - Conditional Average Treatment Effect Summary (Known)} 
\label{tb:caseASummaryKnown}
\end{table}

\begin{figure}[htb]
\centering
\makebox[\textwidth]{
    \includegraphics[scale = 0.35]{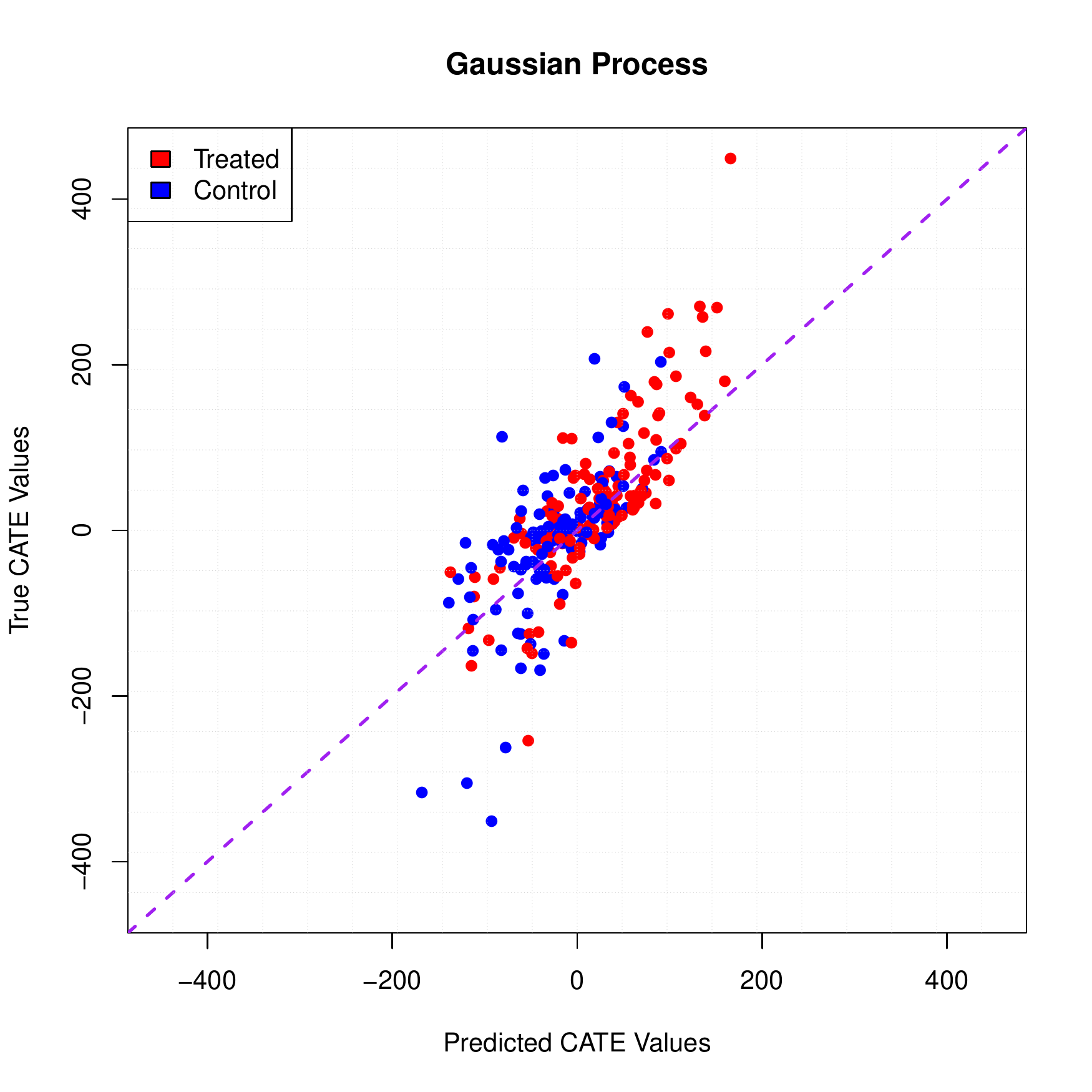}\hspace{-0.4cm}
    \includegraphics[scale = 0.35, page = 1]{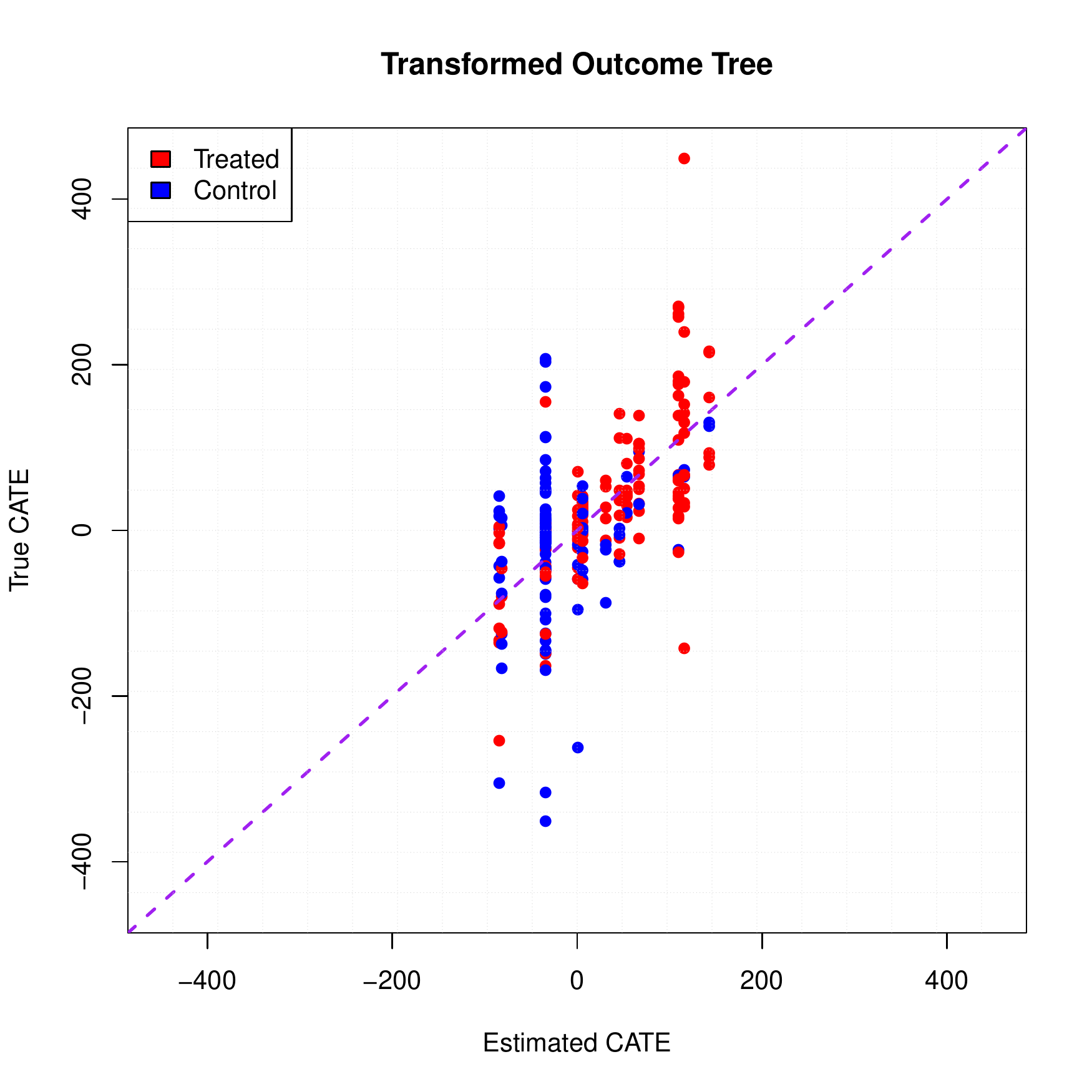}\hspace{-0.4cm}
    \includegraphics[scale = 0.35, page = 1]{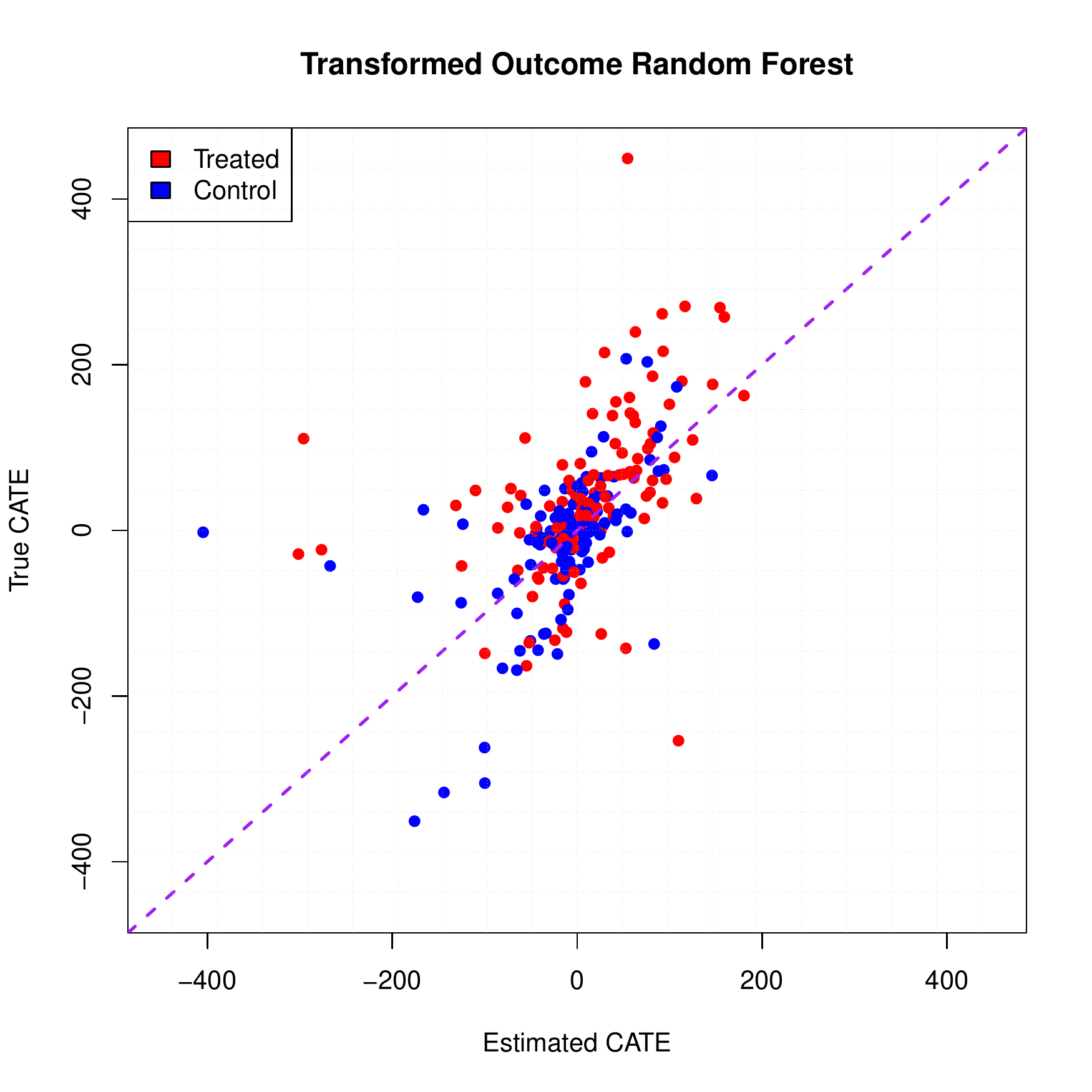}\hspace{-0.4cm}
    } \\ 
    \makebox[\textwidth]{ (a) \hspace{1.8in}  (b) \hspace{1.8in} (c)
    } \\
\makebox[\textwidth]{
    \includegraphics[scale = 0.35, page = 1]{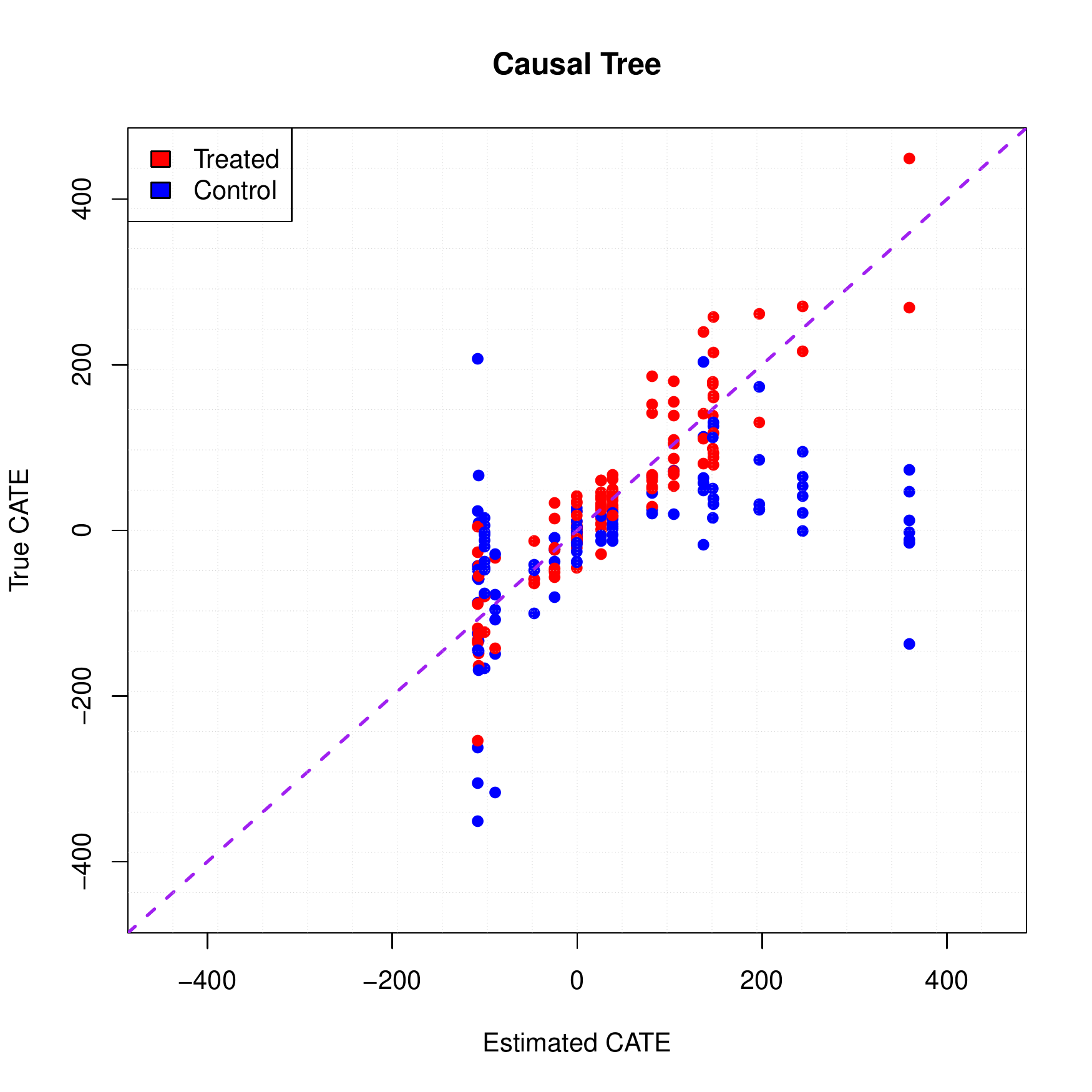}\hspace{-0.4cm}
    \includegraphics[scale = 0.35, page = 1]{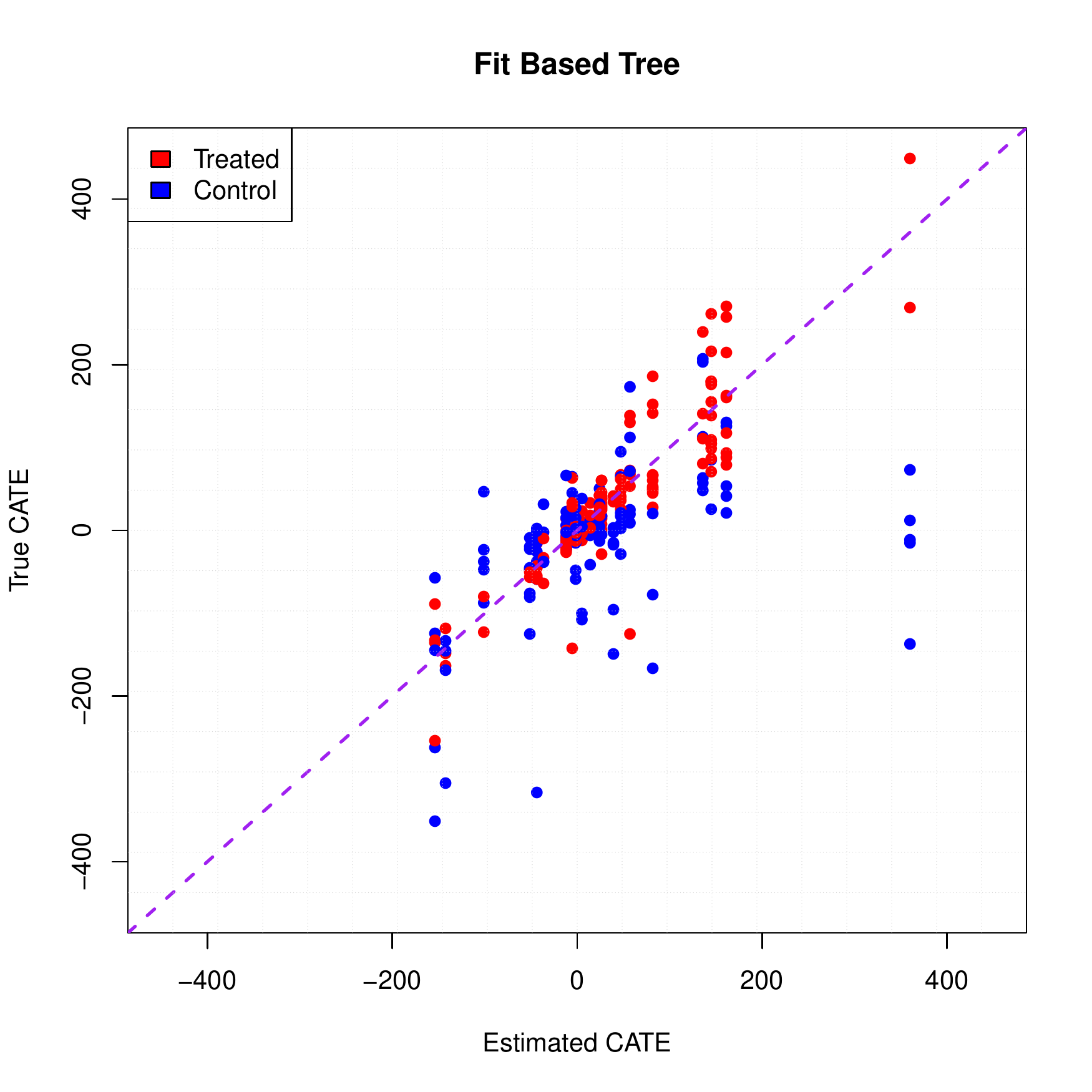}\hspace{-0.4cm}
     \includegraphics[scale = 0.35, page  = 1]{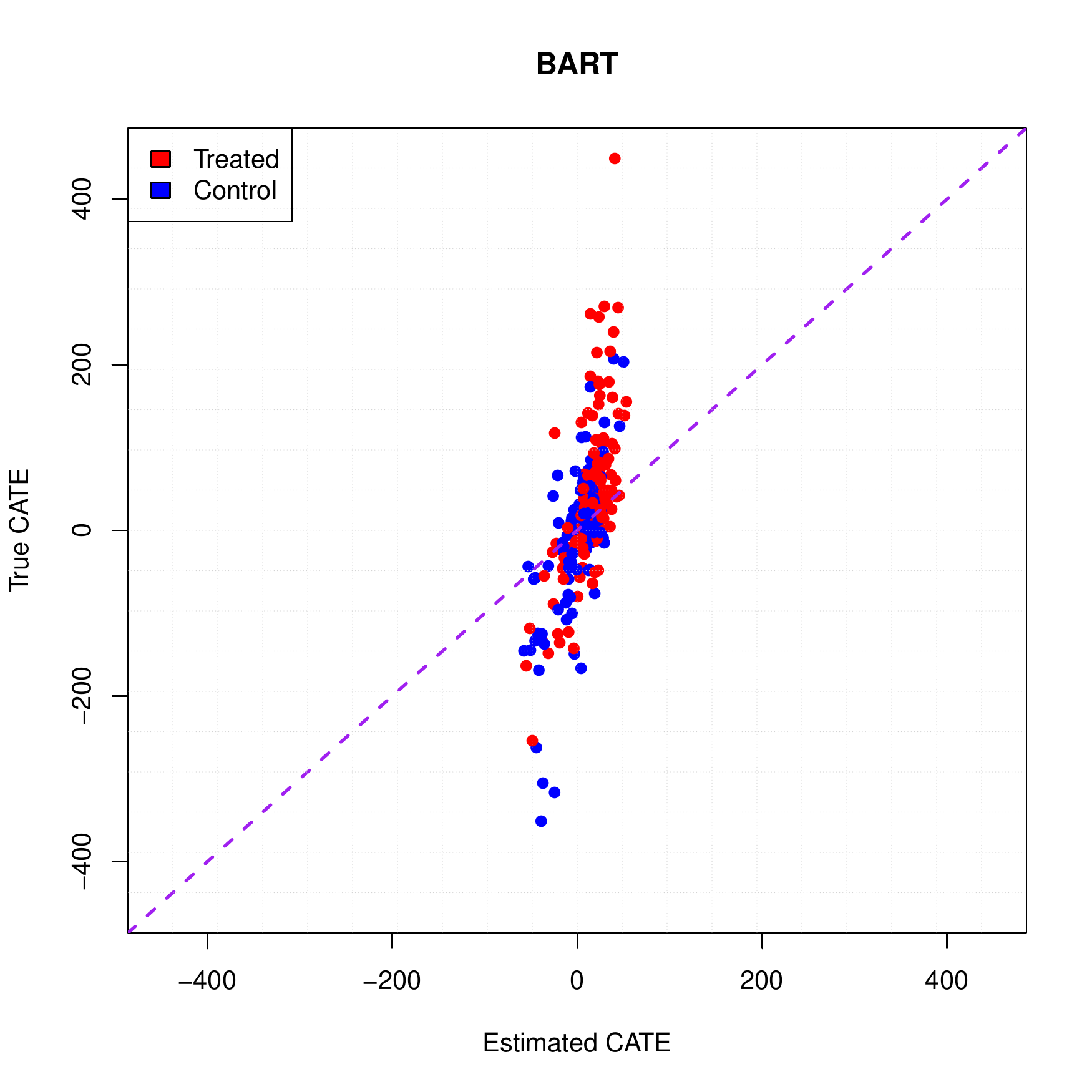}
    
    }\\ 
        \makebox[\textwidth]{ (d) \hspace{1.8in}  (e) \hspace{1.8in} (f) } 
\caption{Comparison of the CATE estimates when the treatment probabilities are unknown for Case A (a) the GP mixture model (b) the transformed outcome regression tree  (c) the transformed outcome random forest (d) the causal tree (e) fit based tree  (f) BART} 
\label{fig:caseAComparisonUnknown}
\end{figure}

\begin{figure}[htb]
\centering
\makebox[\textwidth]{
    \includegraphics[scale = 0.35]{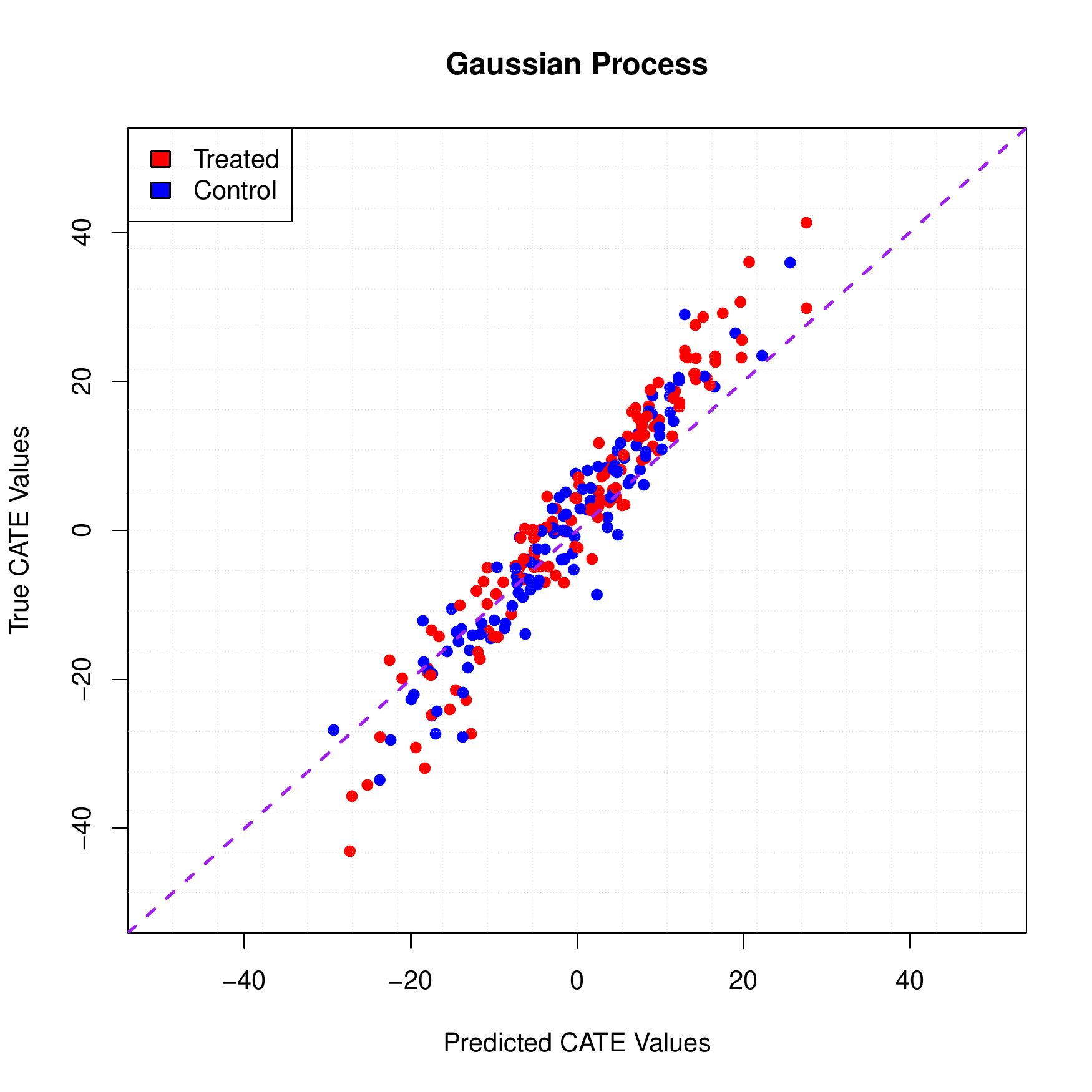}\hspace{-0.4cm}
    \includegraphics[scale = 0.35, page = 2]{"./Results/JMLR/tot_cate_unknown".pdf}\hspace{-0.4cm}
    \includegraphics[scale = 0.35, page = 2]{"./Results/JMLR/rf_cate_unknown".pdf}\hspace{-0.4cm}
    } \\ 
    \makebox[\textwidth]{ (a) \hspace{1.8in}  (b) \hspace{1.8in} (c)
    } \\
\makebox[\textwidth]{
    \includegraphics[scale = 0.35, page = 2]{"./Results/JMLR/ct_cate_unknown".pdf}\hspace{-0.4cm}
    \includegraphics[scale = 0.35, page = 2]{"./Results/JMLR/fit_cate_unknown".pdf}\hspace{-0.4cm}
     \includegraphics[scale = 0.35, page  = 2]{"./Results/JMLR/bart_cate_unknown".pdf}
    
    }\\ 
        \makebox[\textwidth]{ (d) \hspace{1.8in}  (e) \hspace{1.8in} (f) } 
\caption{Comparison of the CATE estimates when the treatment probabilities are unknown for Case B (a) the GP mixture model (b) the transformed outcome regression tree  (c) the transformed outcome random forest (d) the causal tree (e) fit based tree  (f) BART} 
\label{fig:caseBComparisonUnknown}
\end{figure}

We also compare the CATE estimation for both cases when the treatment assignment probabilities are unknown and need to be inferred from the data. The comparison of the point estimation is given in figures \ref{fig:caseAComparisonUnknown} and \ref{fig:caseBComparisonUnknown} respectively for the two cases, with the corresponding summary measurements of fit in tables \ref{tb:caseASummaryUnknown} and \ref{tb:caseBSummaryUnknown}.

\begin{table}[ht]
\centering
\begin{tabular}{rlrrr}
  \hline
 & Model Type & Mean Square Error & Bias &95\% CI Coverage \\ 
  \hline
1 & Gaussian Process Mixture & 50.262 & 3.174 & 0.988 \\ 
  2 & Bayesian Additive Regression Tree & 5.498 & 0.229 & 0.808 \\ 
  3 & Transformed Outcome Tree & 16.421 & 0.202 & 0.900 \\ 
  4 & Fit Based Tree & 15.620 & 0.282 & 0.952 \\ 
  5 & Causal Tree & 21.143 & 0.974 & 0.972 \\ 
  6 & Transformed Outcome Random Forest & 118.745 & -0.582 & 0.816 \\ 
   \hline
\end{tabular}
\caption{Case B - Conditional Average Treatment Effect Summary (Known)} 
\label{tb:caseBSummaryKnown}
\end{table}

For Case A, the performance of the model is far superior in terms of adapting to the heterogeneity, as indicated in figure \ref{fig:caseAComparisonUnknown}(a), in particular compared to the performance of the transformed outcome random forest and BART given in figures  \ref{fig:caseAComparisonUnknown}(c) and \ref{fig:caseAComparisonUnknown}(f). The deterioration in the quality of the estimates from BART is particularly noticeable and can be attributed to the same over-regularization observed before which is even more of a concern since there is additional uncertainty from the learning of the assignment mechanism. Furthermore, while the point-wise coverage of the uncertainty interval is lower relative to the other models, the Gaussian process mixture is the clear winner in terms of the mean square error. The proposed model also outperforms the tree based models (causal and fit based trees as well as transformed outcome trees) in terms of bias (see table \ref{tb:caseASummaryUnknown}) and its point-wise interval coverage is stable relative to BART, which speaks to the models overall robustness despite the added layer of complexity from learning the assignment mechanism.

We see that for Case B, the results of the analysis are similar to when the treatment assignment was known. The performance of the model is comparable in terms of adapting to the heterogeneity relative to the other models, as indicated in figure \ref{fig:caseBComparisonUnknown}(a) -- albeit again with a similar degree of noisiness as earlier. However, we again out-perform transformed outcome random forests in terms of point estimation with lower mean squared error. The only aspect in which the model out performs all the other methods considered is in terms of point-wise interval coverage.

Our conclusion is that the model performs well when there are a large number of covariates present, and the degree of heterogeneity in the treatment effects is high. The flexibility of the mixture of Gaussian processes ensures adaptability, where tree based models fail particularly when there is shared information in the covariates (as is true in Case A) since the prior provides some degree of built-in regularization that is not as excessive as that of BART. However, when the number of covariates is small, the flexibility of the model hurts its overall performance since we observe that our estimates are generally noisier. These limitations of the model are discussed as avenues for future work in the last section of this paper.

\begin{table}[ht]
\centering
\begin{tabular}{rlrrr}
  \hline
 & Model Type & Mean Square Error & Bias &95\% CI Coverage \\ 
  \hline
1 & Gaussian Process Mixture & 3916.562 & 13.207 & 0.780\\ 
  2 & Bayesian Additive Regression Tree & 6754.058 & -5.569 & 0.624 \\ 
  3 & Transformed Outcome Tree & 6289.891 & 7.061 & 0.880 \\ 
  4 & Fit Based Tree & 6154.396 & 15.633 & 0.932 \\ 
  5 & Causal Tree & 8390.039 & 21.923 & 0.968 \\ 
  6 & Transformed Outcome Random Forest & 12124.426 & -21.958 & 0.960 \\ 
   \hline
\end{tabular}
\caption{Case A - Conditional Average Treatment Effect Summary (Unknown)} 
\label{tb:caseASummaryUnknown}
\end{table}

\begin{table}[ht]
\centering
\begin{tabular}{rlrrr}
  \hline
 & Model Type & Mean Square Error & Bias &95\% CI Coverage \\ 
  \hline
1 & Gaussian Process  & 31.517 & 1.898 & 1.000 \\ 
  2 & Bayesian Additive Regression Tree & 6.259 & 0.118 & 0.776\\ 
  3 & Transformed Outcome Tree & 16.421 & 0.202 & 0.892 \\ 
  4 & Fit Based Tree & 15.620 & 0.282 & 0.956\\ 
  5 & Causal Tree & 19.652 & 0.876 & 0.972 \\ 
  6 & Transformed Outcome Random Forest & 115.329 & -0.349 & 0.820  \\ 
   \hline
\end{tabular}
\caption{Case B - Conditional Average Treatment Effect Summary (Unknown)} 
\label{tb:caseBSummaryUnknown}
\end{table}

\subsection{Results on the Italy Survey on Household Income and Wealth (SHIW)}
\label{real}

Our application of the GP mixture model to a real data aimed at the estimation the causal effects of debit card ownership on household spending. A causal analysis of this question was developed in \cite{mercatanti2014debit} using data from the Italy Survey on Household Income and Wealth (SHIW) to estimate the population average treatment effect for the treated (PATT). The SHIW is a biennial, national population representative survey run by Bank of Italy. The subset of the SHIW data we considered consists of  $n = 564$ observations with 385 untreated and 179 treated observations. The outcome variable is the monthly average spending of the household on all consumer goods. The treatment condition is whether the household possesses one and only one debit card, and the control condition is that the household does not possess \emph{any} debit cards.  The covariates we used include: cash inventory held by the household, household income, average interest rate in the province where the household resides, measurement of wealth, and the number of banks in the province in which the household resides. See  \cite{mercatanti2014debit}  for more details about the data. Our analysis of these data will consist of comparing estimates of the ATE and CATE (with respect to household income) of our GP mixture model to the same alternative models as the previous section.

\begin{table}[ht]
\centering
\begin{tabular}{rrrrrr}
  \hline
 Decile &$Mean \quad Income$ & $\widehat{\tau^{CATE}}$ & $\widehat{\tau^{CATE}_{lwr}}$ & $\widehat{\tau^{CATE}_{upr}}$& \\ 
  \hline
1&-1.137 & 0.629 & 0.404 & 0.857 &  \\ 
 2& -0.831 & 0.567 & 0.374 & 0.761 &  \\ 
 3& -0.638 & 0.558 & 0.381 & 0.734 &  \\ 
 4& -0.472 & 0.459 & 0.298 & 0.620 &  \\ 
 5& -0.310 & 0.425 & 0.270 & 0.578 &  \\ 
 6& -0.114 & 0.396 & 0.245 & 0.546 &  \\ 
 7& 0.103 & 0.343 & 0.190 & 0.490 &  \\ 
 8& 0.397 & 0.272 & 0.097 & 0.441 &   \\ 
 9&0.848 & 0.172 & -0.050 & 0.389 &  \\ 
 10& 2.143 & -0.125 & -0.513 & 0.251 & \\ 
   \hline
\end{tabular}
\caption{Conditional average treatment effect with average income by decile} 
\label{tb:cateIncomeRealModel}
\end{table}

\begin{figure}[htb]
\centering
\makebox[\textwidth]{
    \includegraphics[scale = 0.33, page = 1]{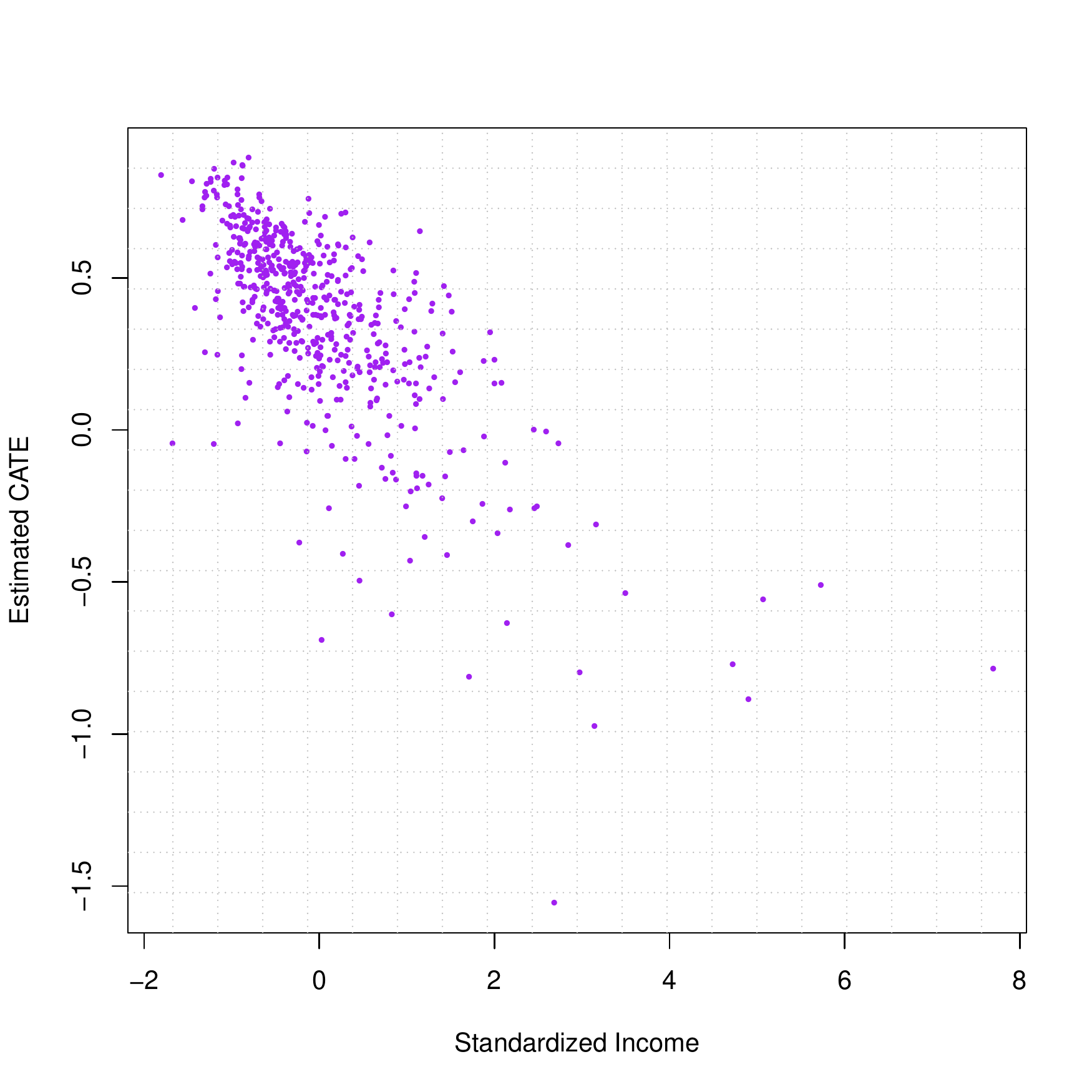}\hspace{-0.2cm} 
    \includegraphics[scale = 0.33, page = 2]{./Results/JMLR/"real_data_model".pdf}\hspace{-0.2cm}
    \includegraphics[scale = 0.33]{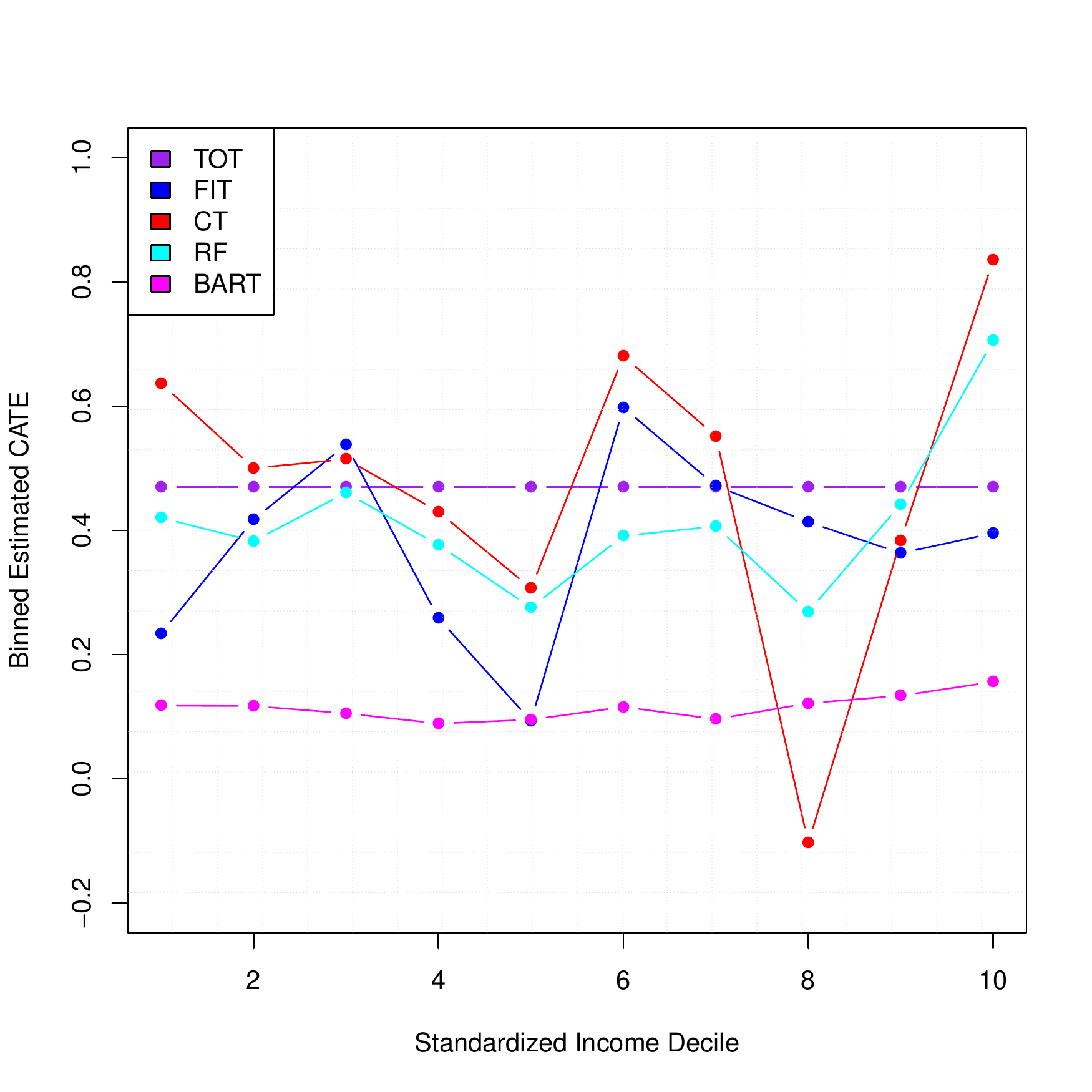}
    }
  \\ \vspace{-.25cm}
    \makebox[\textwidth]{ (a) \hspace{2.2in}  (b) \hspace{2.2in} (c)
    } 
\caption{Estimated CATE (a) against Income (b) binned effects against deciles of income (b) binned effects against deciles of income for comparison}
\label{fig:realComparison}
\end{figure}

We start with a presentation of the CATE under our model against income in \ref{fig:realComparison}(a).The proposed model estimates an overall downward trend in the effect of owning a debit card, i.e. as the level of income increases, the effect of owning a debit card declines. In order to summarize this effect, we consider the CATE for binned deciles of income for the proposed model in figure \ref{fig:realComparison}(b) and the alternative models in figure \ref{fig:realComparison}(c). We find that the proposed model detects a statistically meaningful effect for the first eight deciles of income, and this effect is estimated to decline in size. For the final two deciles, the model concludes that there is no statistically meaningful effect of owning a debit card. These results are summarized in table \ref{tb:cateIncomeRealModel}. By comparison, the inference from the alternative approaches is not quite as clear. BART and transformed outcome trees, detect minimal heterogeneity. With BART, this flattening can be attributed to over-regularization due to the prior, as seen in the simulated data case, while for transformed outcome trees, the axis-parallel splits used to estimate the model are not always suitable for partitioning the covariates. By comparison transformed outcome random forests, transformed outcome trees and causal trees demonstrate the most heterogeneity at the highest two deciles of income. These results are summarized in table \ref{tb:comparisonTable} in  Appendix \ref{app:C}.

In order to be comprehensive and comparable to past work, we have also produced estimates of the average treatment effect in table \ref{tb:ateReal}. The proposed Gaussian process mixture detects a statistically meaningful ATE. This result is consistent with the findings of \cite{mercatanti2014debit}. Furthermore, we also see that the uncertainty interval for the Gaussian process mixture is the tightest of the methods used here, all of which with the exception of BART generate similar inference. This result is consistent with the findings on simulated data presented in the last section since the BART model does not adapt to heterogeneity well in instances where the number of covariates is high with large contributions to the variation in the treatment effects. Again this argues that the GP mixture model may be outperforming the other methods. 

\begin{table}[ht]
\centering
\begin{tabular}{rlrrr}
  \hline
 & Model Type & $\widehat{\tau^{ATE}}$ & $\widehat{\tau^{ATE}_{lwr}}$ & $\widehat{\tau^{ATE}_{upr}}$ \\ 
  \hline
1& Gaussian Process Mixture & 0.369 & 0.220 & 0.518 \\ 
2 & Transformed Outcome Tree & 0.470 & 0.210 & 0.555 \\ 
 3 & Fit Based Tree & 0.378 & 0.214 & 0.608 \\ 
 4 & Causal Tree & 0.475 & 0.360 & 0.939 \\ 
  5& Bayesian Additive Regression Tree & 0.115 & -1.129 & 1.397 \\ 
  6& Transformed Outcome Random Forest & 0.414 & 0.229 & 0.604 \\ 
   \hline
\end{tabular}
\caption{Comparison of average treatment effects.} 
\label{tb:ateReal}
\end{table}

Based on the economic concepts of \emph{income elasticity of demand}, \emph{consumer choice} and \emph{substitution effects} \citep{varian2014intermediate}, the heterogeneity identified at the lowest levels of standardized income is a more sensible result relative to the implication of the other approaches. At the lowest levels of income, economic agents are more likely to substitute debit card use for cash in an effort to maximize spending. The debit cards act as an inflator of perceived financial resources and this effect is expected to diminish as the overall income grows. Therefore, the GP mixture model makes a more convincing case for capturing the true nature of how holding a debit card influences spending. 

\clearpage

\section{Discussion and Future Work}
\label{futurework}
We have proposed a novel non-parametric Bayesian model to estimate heterogeneous treatment effects. Our approach combines the \emph{transformed response variable} framework with  a mixture of Gaussian-processes. The motivation for the GP mixture model was to improve the accuracy of our point estimates as well  as to better quantify uncertainty relative to other models particularly those from the Bayesian non-parametrics literature. We compared the performance of our technique to a single regression tree and random forest model within the TRV framework as well as two conditional mean regression type weighted tree based methods and BART. We used simulation studies to show instances where our approach is a better estimator with respect to both point estimation and  uncertainty quantification.  Furthermore, our approach also has the advantage in that we can address the case where treatment assignment probabilities are unknown within our model; other methods require a two-stage process where another model is required to infer the  treatment assignment probabilities. This tandem estimation provides better insight into the data generating process and also captures uncertainty from all levels of inference.
 
In addition, a Bayesian model of treatment effects with a single likelihood for the design and analysis stages creates concerns of feedback since the TRV depends on the assignment mechanism. We demonstrate that our model is robust to this feedback due to both our prior specification as well as individual covariate adjustment via the Gaussian process covariance functions. However, this raises the theoretical question of whether there is a weaker condition that can be satisfied and still lead to effective inference of treatment effects which is the first area that we aim to explore in future work.

There are several ways we can extend our model to be more robust and flexible. In the context of robustness, the GP prior covariance functions specified impose smoothness assumptions on the treatment effects that may not be realistic in a myriad of applied settings. Relaxing the smoothness and using non-parametric models that have been developed to model dose-response curves may result in richer and more reliable inference. Furthermore, as noted earlier inference using the TRV is sensitive to the probability of receiving the treatment and can create biases and instability when the assignment probability are close to their extremes. While we have addressed instability in the estimation of effects using a correctly specified model and indirectly improved propensity score estimation, we have not directly curbed the susceptibility of the method to extreme weights. The variance of the mixture model is still influenced by the reciprocal of the treatment assignment probability (as is the case generally with IPW estimators). Extending our model to be more insensitive to these extreme cases is vital in application.

Under the theme of model flexibility, we are currently fixing the hyper-parameter values within the kernels of the Gaussian-processes since attempting to learn these from the data creates two problems that we need to carefully study. First, learning these parameters is difficult from a sampling perspective since the target distributions are often extremely multi-modal. A promising avenue for addressing this is the use of a combination of sampling and optimization \citep{levine2001implementations} -- this is particularly important since Bayesian non-parametric methods are known to be sensitive to prior calibration. This is crucial in instances where the degree of heterogeneity in treatment effects is small as we have seen via simulation study. Second, the scalability of Gaussian processes is very limited \citep{johndrow2015approximations} and hence increasing the number of parameters that we are attempting to learn hurts the scalability even more. This broadly summarizes the areas that we will explore in future work.

\begin{appendices}
\section{Proof of Equivalence}\label{app:A}

We now show that the transformation presented in section \ref{former} in expectation recovers the CATE i.e. 
$$\mathbb{E}_{Y}[Y^{*} \mid X = x] = \tau^{CATE}.$$
\noindent
\begin{proof}
First observe that $Y_{i} = Y_{i}(W_{i}) = W_{i}Y_{i}(1) + (1 - W_{i})Y_{i}(0).$

By the definition of the TRV
\begin{eqnarray*}
A = \mathbb{E}_{Y}[Y^{*} \mid X  = x, \mathcal{D}] &=& \mathbb{E}_{Y} \left[\frac{W-e_{i}}{e_{i}(1-e_{i})}Y \mid X = x, \mathcal{D}\right], \\
&=& \frac{1}{e_{i}(1-e_{i})}\left( \mathbb{E}_{Y}[YW \mid X = x, \mathcal{D}] - e_{i} \mathbb{E}_{Y}[Y \mid X=x, \mathcal{D}]\right).
\end{eqnarray*}

\noindent Due to the ignorability of the treatment assignment the following holds
\begin{eqnarray*}
A & = & \frac{1}{e_{i}(1-e_{i})}(e_{i}\mathbb{E}_{Y}[Y \mid W = 1, X=x, \mathcal{D}] - e_{i} \mathbb{E}_{Y}[Y \mid X = x, \mathcal{D}]) \\
&=&\frac{1}{1-e_{i}}\mathbb{E}_{Y}[Y \mid  X = x, W =1,\mathcal{D}] - \frac{1}{1-e_{i}}\mathbb{E}_{Y}[Y \mid X=x, \mathcal{D}].
\end{eqnarray*}
By iterating expectations the following holds:
\begin{eqnarray*}
A &=&\frac{1}{1-e_{i}}\mathbb{E}_{Y}[Y \mid W=1, X=x, \mathcal{D}]  - \frac{1}{1-e_{i}}\mathbb{E}_{W}[\mathbb{E}_{Y}[Y \mid W=1,X=x, \mathcal{D}]], \\
&=&\frac{1}{1-e_{i}}\mathbb{E}_{Y}[Y \mid W = 1,X_{i}=x, \mathcal{D}] -  \frac{e_{i}}{1-e_{i}}\mathbb{E}_{Y}[Y \mid W= 1,X=x,  \mathcal{D}] - \\
& &\mathbb{E}_{Y}[Y \mid W = 0, X=x, \mathcal{D}].
\end{eqnarray*}
Collecting the first two terms provides the desired result
$$A=\mathbb{E}_{Y}[Y \mid W = 1, X= x, \mathcal{D}] - \mathbb{E}_{Y}[Y \mid W =0, X=x, \mathcal{D}].$$
\end{proof}

\section{Derivation of Model}\label{app:B}
The derivation of the model presented in the paper begins with the transformation of interest given as follows, with $Y_{i}$ denoting the observed response, $W_{i}$ the assigned treatment and $e_{i} = P(W_{i} = 1)$
\begin{equation*}
Y_{i}^{*} = \frac{W_{i} - e_{i}}{e_{i}(1-e_{i})}Y_{i}.
\end{equation*}
In addition, we define the two regression functions for the outcome, one under the treatment and one under the control,
\begin{align*}
(Y_{i}|W_{i} = 0) = f_{0}(X_{i})+\epsilon_{i}(0),\\
(Y_{i}|W_{i} = 1) = f_{1}(X_{i})+\epsilon_{i}(1).\\
\end{align*}
Using the transformation, and substituting the regression functions under the two cases i.e. when $W_{i} = 1$ and when $W_{i} = 0$ and assuming further that $\epsilon(1), \epsilon(0) \stackrel{IID}{\sim}\mathrm{N}(0, \sigma^{2})$, we can define with probability $e_{i}$,

\begin{align*}
(Y_{i}^{*}|W_{i} = 1) &= \frac{f_{1}(X_{i}) - e_{i}f_{1}(X_{i})+e_{i}f_{0}(X_{i})}{e_{i}} + f_{1}(X_{i})-f_{0}(X_{i}) + \frac{\epsilon_{i}(1)}{e_{i}},\\
&= f_{1}(X_{i})-f_{0}(X_{i}) + (1-e_{i})\bigg(\frac{ f_{1}(X_{i})}{e_{i}}+\frac{f_{0}(X_{i})}{1-e_{i}} \bigg)+\frac{\epsilon_{i}(1)}{e_{i}},\\
&= g(X_{i})+ (1-e_{i})h(X_{i})+\frac{\epsilon_{i}(1)}{e_{i}}.
\end{align*}

and similarly, with probability $1-e_{i}$ that,

\begin{align*}
(Y_{i}^{*}|W_{i} = 0) &= \frac{-(1-e_{i})f_{1}(X_{i}) +(1- e_{i})f_{0}(X_{i})-f_{0}(X_{i})}{e_{i}} + f_{1}(X_{i})-f_{0}(X_{i}) - \frac{\epsilon_{i}(0)}{1-e_{i}},\\
&= f_{1}(X_{i})-f_{0}(X_{i}) + (-e_{i})\bigg(\frac{ f_{1}(X_{i})}{e_{i}}+\frac{f_{0}(X_{i})}{1-e_{i}} \bigg)-\frac{\epsilon_{i}(0)}{1-e_{i}},\\
&= g(X_{i})+ (-e_{i})h(X_{i})+\frac{\epsilon_{i}(0)}{e_{i}-1}.
\end{align*}

This yields the mixture model model that we have presented in the paper,
\begin{align*}
Y_{i}^{*} &= g(X_{i}) + \varepsilon_{i},\\
\varepsilon_{i}\sim(e_{i})\mathrm{Normal}((1-e_{i})h(X_{i}), \frac{\sigma^{2}}{e_{i}^{2}})&+(1-e_{i})\mathrm{Normal}(-e_{i}h(X_{i}), \frac{\sigma^{2}}{(1-e_{i})^{2}}).
\end{align*}

\section{Comparison of SHIW Data}
\label{app:C}
This section presents comparative analysis using various methods for the CATE estimation for the SHIW data using the Gaussian process mixture in section \ref{real}. Point estimates of the CATE along with $95\%$ uncertainty intervals for each decile of income, along with the average value of income in that decile are presented in table \ref{tb:comparisonTable}.

\section{Sampling Algorithms for Model Specifications}
\label{app:D}

\paragraph{Algorithm for inference with known assignment probabilities:}
For the full conditional distributions specified in \eqref{eq:fcg} we can run the following Gibbs sampling procedure to generate a sequence $(\mathbf{g}^{(j)}, \mathbf{h}^{(j)}, \sigma^{((j)})_{j=1}^K$ as follows,

\begin{enumerate}
\item[a)] Initialize ${\mathbf h}^{(0)}$, $\sigma^{(0)}$, and ${\mathbf g}^{(0)}$;
 \item[b)] For $j= 1,...,K$
\begin{enumerate}
\item[1)] $\mathbf{g}^{(j)} \sim  \pi(\mathbf{g} \mid \mathbf{h}^{(j-1)}, \sigma^{(j-1)}, \mathcal{D})$;
\item[2)] $\mathbf{h}^{(j)} \sim  \pi(\mathbf{h} \mid \mathbf{g}^{(j)}, \sigma^{(j-1)}, \mathcal{D})$ ;
\item[2)] $ \sigma^{(j)} \sim \pi(\sigma \mid \mathbf{h}^{(j)}, \mathbf{g}^{(j)}, \mathcal{D}).$
\end{enumerate} 
\end{enumerate}

Given the sequence $(\mathbf{g}^{(j)}, \mathbf{h}^{(j)}, \sigma^{(j)})_{j=1}^K$ we discard an initial $K_0$ of the samples to address burn-in of the chain and we thin the remaining samples by a small factor $\gamma$ to obtain independent samples from the joint posterior distribution in section \ref{simspec} and in equation \eqref{eq:jd1}. We will specify the burn-in and thinning settings whenever we discuss applications of the method.

\paragraph{Algorithm for inference with unknown assignment probabilities:} For the full posterior stated in equation \eqref{post3} a standard Gibbs sampling procedure of the type specified above cannot be used for sampling the treatment assignment probabilities. We use a n\"aive approach to sampling the assignment probabilities in addition to the other model parameters with an additional Metropolis-within-Gibbs step. This results in the following procedure:
\begin{enumerate}
\item[a)] Initialize ${\mathbf h}^{(0)}$, $\sigma^{(0)}$, ${\mathbf g}^{(0)}$, and $\boldsymbol{\beta}^{(0)}$. Use $\boldsymbol{\beta}^{(0)}$to compute $\boldsymbol{e}^{(0)}$;
\item[b)] Compute $\mathbf{Y}^*$ from the initial $\boldsymbol{e}^{(0)}$ and data;
 \item[c)] For $j= 1,...,K$
\begin{enumerate}
\item[1)] $\mathbf{g}^{(j)} \sim  \pi(\mathbf{g} \mid \mathbf{h}^{(j-1)}, \sigma^{(j-1)}, \boldsymbol{e}^{(j-1)}, \mathbf{Y}^*, \mathcal{D})$;
\item[2)] $\mathbf{h}^{(j)} \sim  \pi(\mathbf{h} \mid \mathbf{g}^{(j)}, \sigma^{(j-1)}, \boldsymbol{e}^{(j-1)}, \mathbf{Y}^*, \mathcal{D})$ ;
\item[3)] $ \sigma^{(j)} \sim \pi(\sigma \mid \mathbf{h}^{(j)}, \mathbf{g}^{(j)}, \boldsymbol{e}^{(j-1)}, \mathbf{Y}^*, \mathcal{D})$;
\item[4)] Use Metropolis-Hastings step to sample $\boldsymbol{\beta}^{(j)}$;
\item[5)] Compute $\boldsymbol{e}^{(j)}$ from $\boldsymbol{\beta}^{(j)}$ and data;
\item[6)] Compute $\mathbf{Y}^*$ from $\boldsymbol{e}^{(j)}$ and data.
\end{enumerate}
\end{enumerate}
Therefore using the steps in the algorithm above we simulate a sequence $(\mathbf{g}^{(j)}, \mathbf{h}^{(j)}, \sigma^{(j)}, \pmb{\beta}^{(j)},$ $\pmb{e}^{(j)},\mathbf{Y}^{*(j)})_{j=1}^K$ akin to earlier with burn-in and thinning considerations that reflects draws from the joint distribution in section \ref{compspec1} in \eqref{post3}.

\end{appendices}
\clearpage
\begin{sidewaystable}
\begin{center}
\tiny
\begin{tabular}{rrrrrrrrrrrrrrrrrr}
  \hline
 Decile & $Mean \quad Income$ & $tot$ & $fit$ & $ct$ & $BART$ & $RF$ & $lwr_{tot}$ & $upr_{tot}$ & $lwr_{fit}$ & $upr_{fit}$ &$lwr_{ct}$ & $upr_{ct}$ & $lwr_{BART}$ & $upr_{BART}$ & $lwr_{RF}$ & $upr_{RF}$ \\ 
  \hline
1 & -1.137 & 0.470 & 0.234 & 0.637 & 0.118 & 0.420 & 0.089 & 0.678 & 0.087 & 0.741 & 0.202 & 1.103 & -1.166 & 1.394 & 0.082 & 0.772 \\ 
2  & -0.831 & 0.470 & 0.417 & 0.500 & 0.117 & 0.383 & 0.087 & 0.672 & 0.096 & 0.730 & 0.191 & 1.117 & -1.211 & 1.397 & 0.083 & 0.770 \\ 
3  & -0.638 & 0.470 & 0.538 & 0.515 & 0.105 & 0.461 & 0.067 & 0.679 & 0.083 & 0.733 & 0.183 & 1.094 & -1.204 & 1.398 & 0.082 & 0.772 \\ 
4 & -0.472 & 0.470 & 0.258 & 0.430 & 0.089 & 0.376 & 0.076 & 0.676 & 0.094 & 0.725 & 0.193 & 1.105 & -1.224 & 1.393 & 0.085 & 0.744 \\ 
5   & -0.310 & 0.470 & 0.093 & 0.307 & 0.095 & 0.276 & 0.077 & 0.670 & 0.082 & 0.740 & 0.198 & 1.112 & -1.191 & 1.398 & 0.075 & 0.766 \\ 
6  & -0.114 & 0.470 & 0.598 & 0.681 & 0.116 & 0.391 & 0.074 & 0.676 & 0.097 & 0.732 & 0.204 & 1.103 & -1.107 & 1.397 & 0.083 & 0.772 \\ 
7   & 0.103 & 0.470 & 0.471 & 0.552 & 0.096 & 0.407 & 0.081 & 0.660 & 0.086 & 0.748 & 0.212 & 1.112 & -1.136 & 1.370 & 0.087 & 0.760 \\ 
8   & 0.397 & 0.470 & 0.414 & -0.103 & 0.122 & 0.269 & 0.086 & 0.682 & 0.092 & 0.744 & 0.192 & 1.104 & -1.120 & 1.385 & 0.082 & 0.760 \\ 
9   & 0.848 & 0.470 & 0.363 & 0.384 & 0.134 & 0.442 & 0.082 & 0.667 & 0.094 & 0.744 & 0.178 & 1.108 & -1.119 & 1.400 & 0.087 & 0.776 \\ 
10   & 2.143 & 0.470 & 0.396 & 0.835 & 0.156 & 0.706 & 0.075 & 0.673 & 0.107 & 0.735 & 0.190 & 1.108 & -1.068 & 1.406 & 0.081 & 0.760 \\ 
   \hline
\end{tabular}
\end{center}
\caption{Comparison of conditional average treatment effects by decile of standardized income, along with 95\% uncertainty intervals using alternative models.}
\label{tb:comparisonTable}
\end{sidewaystable}

\clearpage
\section{Acknowledgements}
The authors gratefully acknowledge the support of  Andrea Mercatanti (Department of Statistics, Bank of Italy) for providing data for this paper and sincerely thank Elizabeth Lorenzi (Duke University) for providing insightful commentary and expertise on the topic of causal inference.

\clearpage
\bibliography{refs.bib}

\begin{thebibliography}{41}
\providecommand{\natexlab}[1]{#1}
\providecommand{\url}[1]{\texttt{#1}}
\expandafter\ifx\csname urlstyle\endcsname\relax
  \providecommand{\doi}[1]{doi: #1}\else
  \providecommand{\doi}{doi: \begingroup \urlstyle{rm}\Url}\fi

\bibitem[Athey(2015)]{Athey:2015:MLC:2783258.2785466}
Susan Athey.
\newblock {M}achine {L}earning and {C}ausal {I}nference for {P}olicy
  {E}valuation.
\newblock In \emph{Proceedings of the 21th {ACM SIGKDD} {I}nternational
  {C}onference on {K}nowledge {D}iscovery and {D}ata {M}ining}, KDD '15, pages
  5--6, New York, NY, USA, 2015. ACM.
\newblock ISBN 978-1-4503-3664-2.
\newblock \doi{10.1145/2783258.2785466}.
\newblock URL \url{http://doi.acm.org/10.1145/2783258.2785466}.

\bibitem[{A}they and {I}mbens(2016)]{athey2016recursive}
{S}usan {A}they and {G}uido {I}mbens.
\newblock {R}ecursive {P}artitioning for {H}eterogeneous {C}ausal {E}ffects.
\newblock \emph{{P}roceedings of the {N}ational {A}cademy of {S}ciences},
  113\penalty0 (27):\penalty0 7353--7360, 2016.

\bibitem[Athey and Imbens(2015)]{athey2015machine}
Susan Athey and Guido~W Imbens.
\newblock {M}achine {L}earning {M}ethods for {E}stimating {H}eterogeneous
  {C}ausal {E}ffects.
\newblock \emph{{S}tat}, 1050:\penalty0 5, 2015.

\bibitem[Beygelzimer and Langford(2009)]{beygelzimer2009offset}
Alina Beygelzimer and John Langford.
\newblock {T}he {O}ffset {T}ree for {L}earning with {P}artial {L}abels.
\newblock In \emph{{P}roceedings of the 15th {ACM SIGKDD} {I}nternational
  {C}onference on {K}nowledge {D}iscovery and {D}ata {M}ining}, pages 129--138.
  ACM, 2009.

\bibitem[Breiman(1984)]{breiman1984classification}
Leo Breiman.
\newblock {C}lassification and {R}egression {T}rees.
\newblock 1984.

\bibitem[Breiman(2001)]{breiman2001random}
Leo Breiman.
\newblock {R}andom {F}orests.
\newblock \emph{{M}achine {L}earning}, 45\penalty0 (1):\penalty0 5--32, 2001.

\bibitem[Chipman et~al.(2010)Chipman, George, McCulloch,
  et~al.]{chipman2010bart}
Hugh~A Chipman, Edward~I George, Robert~E McCulloch, et~al.
\newblock {BART}: {B}ayesian {A}dditive {R}egression {T}rees.
\newblock \emph{{T}he {A}nnals of {A}pplied {S}tatistics}, 4\penalty0
  (1):\penalty0 266--298, 2010.

\bibitem[Ding and Li(2017)]{ding2017causal}
Peng Ding and Fan Li.
\newblock {C}ausal {I}nference: {A} {M}issing {D}ata {P}erspective.
\newblock \emph{arXiv preprint arXiv:1712.06170}, 2017.

\bibitem[Dud{\'\i}k et~al.(2011)Dud{\'\i}k, Langford, and Li]{dudik2011doubly}
Miroslav Dud{\'\i}k, John Langford, and Lihong Li.
\newblock {D}oubly {R}obust {P}olicy {E}valuation and {L}earning.
\newblock \emph{arXiv preprint arXiv:1103.4601}, 2011.

\bibitem[Foster et~al.(2011)Foster, Taylor, and Ruberg]{foster2011subgroup}
Jared~C Foster, Jeremy~MG Taylor, and Stephen~J Ruberg.
\newblock {S}ubgroup {I}dentification {F}rom {R}andomized {C}linical {T}rial
  {D}ata.
\newblock \emph{Statistics in medicine}, 30\penalty0 (24):\penalty0 2867--2880,
  2011.

\bibitem[Green and Kern(2012)]{green2012modeling}
Donald~P Green and Holger~L Kern.
\newblock {M}odeling {H}eterogeneous {T}reatment {E}ffects in {S}urvey
  {E}xperiments with {B}ayesian {A}additive {R}egression {T}rees.
\newblock \emph{{P}ublic {O}pinion {Q}uarterly}, 76\penalty0 (3):\penalty0
  491--511, 2012.

\bibitem[Hahn et~al.(2017)Hahn, Murray, and Carvalho]{hahn2017bayesian}
P~Richard Hahn, Jared~S Murray, and Carlos~M Carvalho.
\newblock {B}ayesian {R}egression {T}ree {M}odels for {C}ausal inference:
  {R}egularization, {C}onfounding, and {H}eterogeneous {E}ffects.
\newblock 2017.

\bibitem[{H}ahn et~al.(2018){H}ahn, {C}arvalho, {P}uelz, {H}e, and
  {O}thers]{hahn2018regularization}
{P}~{R}ichard {H}ahn, {C}arlos~{M} {C}arvalho, {D}avid {P}uelz, {J}ingyu {H}e,
  and {O}thers.
\newblock {R}egularization and {C}onfounding in {L}inear {R}egression for
  {T}reatment {E}ffect {E}stimation.
\newblock \emph{{B}ayesian {A}nalysis}, 13\penalty0 (1):\penalty0 163--182,
  2018.

\bibitem[Heckman et~al.(2006)Heckman, Urzua, and
  Vytlacil]{heckman2006understanding}
James~J Heckman, Sergio Urzua, and Edward Vytlacil.
\newblock {U}nderstanding {I}nstrumental {V}ariables in {M}odels with
  {E}ssential {H}eterogeneity.
\newblock \emph{{T}he Review of {E}conomics and {S}tatistics}, 88\penalty0
  (3):\penalty0 389--432, 2006.

\bibitem[Hill(2011)]{hill2011bayesian}
Jennifer~L Hill.
\newblock {B}ayesian {N}onparametric {M}odeling for {C}ausal {I}nference.
\newblock \emph{{J}ournal of {C}omputational and {G}raphical {S}tatistics},
  20\penalty0 (1):\penalty0 217--240, 2011.

\bibitem[{H}irano et~al.(2003){H}irano, {I}mbens, and
  {R}idder]{hirano2003efficient}
{K}eisuke {H}irano, {G}uido~{W} {I}mbens, and {G}eert {R}idder.
\newblock {E}fficient {E}stimation of {A}verage {T}reatment {E}ffects {U}sing
  the {E}stimated {P}ropensity {S}core.
\newblock \emph{{E}conometrica}, 71\penalty0 (4):\penalty0 1161--1189, 2003.

\bibitem[Imbens and Rubin(2015)]{imbens2015causal}
Guido~W Imbens and Donald~B Rubin.
\newblock \emph{{C}ausal {I}nference in {S}tatistics, {S}ocial, and
  {B}iomedical {S}ciences}.
\newblock Cambridge University Press, 2015.

\bibitem[Johndrow et~al.(2015)Johndrow, Mattingly, Mukherjee, and
  Dunson]{johndrow2015approximations}
James~E Johndrow, Jonathan~C Mattingly, Sayan Mukherjee, and David Dunson.
\newblock {A}pproximations of {M}arkov {C}hains and {B}ayesian {I}nference.
\newblock \emph{arXiv preprint arXiv:1508.03387}, 2015.

\bibitem[{L}evine and {C}asella(2001)]{levine2001implementations}
{R}ichard~{A} {L}evine and {G}eorge {C}asella.
\newblock {I}mplementations of the {M}onte {C}arlo {EM } {A}lgorithm.
\newblock \emph{{J}ournal of {C}omputational and {G}raphical {S}tatistics},
  10\penalty0 (3):\penalty0 422--439, 2001.

\bibitem[Liaw and Wiener(2002)]{rpart2002}
Andy Liaw and Matthew Wiener.
\newblock {C}lassification and {R}egression by random{F}orest.
\newblock \emph{R News}, 2\penalty0 (3):\penalty0 18--22, 2002.
\newblock URL \url{https://CRAN.R-project.org/doc/Rnews/}.

\bibitem[Mercatanti et~al.(2014)Mercatanti, Li, et~al.]{mercatanti2014debit}
Andrea Mercatanti, Fan Li, et~al.
\newblock {D}o {D}ebit {C}ards {I}ncrease {H}ousehold {S}pending? {E}vidence
  from a {S}emiparametric {C}ausal {A}nalysis of a {S}urvey.
\newblock \emph{The {A}nnals of {A}pplied {S}tatistics}, 8\penalty0
  (4):\penalty0 2485--2508, 2014.

\bibitem[Pearl et~al.(2009)]{pearl2009causal}
Judea Pearl et~al.
\newblock {C}ausal {I}nference in {S}tatistics: an {O}verview.
\newblock \emph{{S}tatistics {S}urveys}, 3:\penalty0 96--146, 2009.

\bibitem[Powers et~al.(2017)Powers, Qian, Jung, Schuler, Shah, Hastie, and
  Tibshirani]{powers2017some}
Scott Powers, Junyang Qian, Kenneth Jung, Alejandro Schuler, Nigam~H Shah,
  Trevor Hastie, and Robert Tibshirani.
\newblock {S}ome {M}ethods for {H}eterogeneous {T}reatment {E}ffect
  {E}stimation in {H}igh-{D}imensions.
\newblock \emph{arXiv preprint arXiv:1707.00102}, 2017.

\bibitem[Pratola et~al.(2016)]{pratola2016efficient}
Matthew~T Pratola et~al.
\newblock Efficient metropolis--hastings proposal mechanisms for bayesian
  regression tree models.
\newblock \emph{Bayesian analysis}, 11\penalty0 (3):\penalty0 885--911, 2016.

\bibitem[Rasmussen(2000)]{rasmussen2000infinite}
Carl~Edward Rasmussen.
\newblock The {I}nfinite {G}aussian {M}ixture {M}odel.
\newblock In \emph{{A}dvances in {N}eural {I}nformation {P}rocessing
  {S}ystems}, pages 554--560, 2000.

\bibitem[Rasmussen(2006)]{rasmussen2006Gaussian}
Carl~Edward Rasmussen.
\newblock Gaussian {P}rocesses for {M}achine {L}earning.
\newblock 2006.

\bibitem[{R}obert {M}c{C}ulloch et~al.(2018){R}obert {M}c{C}ulloch, {R}odney
  {S}parapani, {R}obert Gramacy, {C}harles {S}panbauer, and {M}atthew
  {P}ratola]{bart2018}
{R}obert {M}c{C}ulloch, {R}odney {S}parapani, {R}obert Gramacy, {C}harles
  {S}panbauer, and {M}atthew {P}ratola.
\newblock \emph{{BART}: {B}ayesian {A}dditive {R}egression {T}rees}, 2018.
\newblock URL \url{https://CRAN.R-project.org/package=BART}.
\newblock R package version 1.9.

\bibitem[{R}osenbaum and {R}ubin(1982)]{rosenbaum1982assessing}
{P}aul~{R} {R}osenbaum and {D}onald~{B} {R}ubin.
\newblock \emph{{A}ssessing {S}ensitivity to an {U}nobserved {B}inary
  {C}ovariate in an {O}bservational {S}tudy with {B}inary {O}utcome}.
\newblock {W}isconsin {C}linical {C}ancer {C}enter, {B}iostatistics, 1982.

\bibitem[Rubin(1978)]{rubin1978bayesian}
Donald~B Rubin.
\newblock Bayesian {I}nference for {C}ausal {E}ffects: The {R}ole of
  {R}andomization.
\newblock \emph{{T}he {A}nnals of {S}tatistics}, pages 34--58, 1978.

\bibitem[Rubin(2005)]{rubin2005causal}
Donald~B Rubin.
\newblock {C}ausal {I}nference {U}sing {P}otential {O}utcomes: {D}esign,
  {M}odeling, {D}ecisions.
\newblock \emph{{J}ournal of the {A}merican {S}tatistical {A}ssociation},
  100\penalty0 (469):\penalty0 322--331, 2005.

\bibitem[Rubin and Thomas(1996)]{rubin1996matching}
Donald~B Rubin and Neal Thomas.
\newblock {M}atching {U}sing {E}stimated {P}ropensity {S}cores: {R}elating
  {T}heory to {P}ractice.
\newblock \emph{{B}iometrics}, pages 249--264, 1996.

\bibitem[Singh(2016)]{singh2016gaussian}
Susheela Singh.
\newblock {G}aussian {P}rocess {R}egression.
\newblock 2016.

\bibitem[{S}usan {A}they et~al.(2016){S}usan {A}they, {G}uido {I}mbens, and
  {Y}anyang {K}ong]{causal2016tree}
{S}usan {A}they, {G}uido {I}mbens, and {Y}anyang {K}ong.
\newblock \emph{causal{T}ree: {R}ecursive {P}artitioning {C}ausal {T}rees},
  2016.
\newblock R package version 0.0.

\bibitem[{T}erry {T}herneau and {B}eth {A}tkinson(2018)]{rf2018}
{T}erry {T}herneau and {B}eth {A}tkinson.
\newblock \emph{rpart: {R}ecursive {P}artitioning and {R}egression {T}rees},
  2018.
\newblock URL \url{https://CRAN.R-project.org/package=rpart}.
\newblock R package version 4.1-13.

\bibitem[Tibshirani(1996)]{tibshirani1996regression}
Robert Tibshirani.
\newblock {R}egression {S}hrinkage and {S}election {V}ia the {L}asso.
\newblock \emph{{J}ournal of the {R}oyal {S}tatistical {S}ociety. Series B
  (Methodological)}, pages 267--288, 1996.

\bibitem[Varian(2014)]{varian2014intermediate}
Hal~R Varian.
\newblock \emph{{I}ntermediate {M}icroeconomics: {A} {M}odern {A}pproach:
  {N}inth {I}nternational {S}tudent {E}dition}.
\newblock WW Norton \& Company, 2014.

\bibitem[Wager and Athey(2017)]{wager2017estimation}
Stefan Wager and Susan Athey.
\newblock {E}stimation and {I}nference of {H}eterogeneous {T}reatment {E}ffects
  {U}sing {R}andom {F}orests.
\newblock \emph{{J}ournal of the {A}merican {S}tatistical {A}ssociation},
  \penalty0 (just-accepted), 2017.

\bibitem[Wager et~al.(2014)Wager, Hastie, and Efron]{wager2014confidence}
Stefan Wager, Trevor Hastie, and Bradley Efron.
\newblock {C}onfidence {I}ntervals for {R}andom {F}orests: {T}he {J}ackknife
  and the {I}nfinitesimal {J}ackknife.
\newblock \emph{The {J}ournal of {M}achine {L}earning {R}esearch}, 15\penalty0
  (1):\penalty0 1625--1651, 2014.

\bibitem[{W}endling et~al.(2018){W}endling, {J}ung, {C}allahan, {S}chuler,
  {S}hah, and {G}allego]{wendling2018comparing}
{T} {W}endling, {K} {J}ung, {A} {C}allahan, {A} {S}chuler, {NH} {S}hah, and {B}
  {G}allego.
\newblock {C}omparing {M}ethods for {E}stimation of {H}eterogeneous {T}reatment
  {E}ffects using {O}bservational {D}ata from {H}ealth {C}are {D}atabases.
\newblock \emph{{S}tatistics in {M}edicine}, 2018.

\bibitem[Xie et~al.(2012)Xie, Brand, and Jann]{xie2012estimating}
Yu~Xie, Jennie~E Brand, and Ben Jann.
\newblock {E}stimating {H}eterogeneous {T}reatment {E}ffects with
  {O}bservational data.
\newblock \emph{{S}ociological {M}ethodology}, 42\penalty0 (1):\penalty0
  314--347, 2012.

\bibitem[{Z}igler et~al.(2013){Z}igler, {W}atts, {Y}eh, {W}ang, {C}oull, and
  {D}ominici]{zigler2013model}
{C}orwin~{M} {Z}igler, {K}rista {W}atts, {R}obert~{W} {Y}eh, {Y}un {W}ang,
  {B}rent~{A} {C}oull, and {F}rancesca {D}ominici.
\newblock {M}odel {F}eedback in {B}ayesian {P}ropensity {S}core {E}stimation.
\newblock \emph{{B}iometrics}, 69\penalty0 (1):\penalty0 263--273, 2013.

\end{thebibliography}

\end{document}